\documentclass[acmlarge,screen]{acmart}

\def\BibTeX{{\rm B\kern-.05em{\sc i\kern-.025em b}\kern-.08emT\kern-.1667em\lower.7ex\hbox{E}\kern-.125emX}}

\usepackage{acronym}

\newacro{CACHET}{Copenhagen Center for Health Technology} 
\newacro{EMA}{Ecological Momentary Assessment} 
\newacro{PRO}{Patient Reported Outcome} 
\newacro{API}{application programming interface} 
\newacro{GDPR}{General Data Protection Regulation} 
\newacro{HL7}{Health Level 7} 
\newacro{FHIR}{Fast Healthcare Interoperability Resources} 
\newacro{BTLE}{Bluetooth Low Energy} 
\newacro{BA}{Behavioral Activation} 
\newacro{CBT}{Cognitive Behavioral Therapy} 
\newacro{PI}{Personal Informatics} 
\newacro{Xamarin}{Xamarin} 
\newacro{PCD}{Patient-Clinician-Designer framework}
\newacro{PHQ-8}{Patient Health Questionnaire} 
\newacro{WHO-5}{WHO Well-being Index} 
\newacro{OMH}{Open mHealth} 
\newacro{EU}{European Union} 
\newacro{AWS}{Amazon Web Services} 
\newacro{PSSUQ}{Post-Study System Usability Questionnaire}
\newacro{SUS}{System Usability Scale}
\newacro{mHealth}{mobile health} 
\newacro{IPSRT}{Interpersonal and Social Rhythm Therapy} 
\newacro{DBT}{Dialectical Behavior Therapy} 
\newacro{MOSS}{Mobile Sensing and Support} 
\newacro{RCT}{randomized controlled trial} 
\newacro{UD}{unipolar disorder}
\newacro{BD}{bipolar disorder}

\newacro{ECG}{electrocardiography} 
\newacro{CGM}{continuous glucose monitoring} 
\newacro{CARP}{CACHET Research Platform} 
\newacro{CAMS}{CARP Mobile Sensing} 
\newacro{API}{application programming interface} 
\newacro{APK}{Android Application Package}
\newacro{OS}{operating system} 
\newacro{BLoC}{Business Logic Component} 
\newacro{HR}{heart rate} 
\newacro{HRV}{heart rate variability} 
\newacro{TAM}{Technology Acceptance Model} 
\newacro{UTAUT}{Unified Theory of Acceptance and Use of Technology} 
\newacro{UI}{user interface} 
\newacro{IMU}{Inertial Measurement Unit} 
\newacro{PPG}{photoplethysmogram} 
\newacro{US}{United States} 
\newacro{CVD}{cardiovascular disease} 

\usepackage{hyperref}
\usepackage{listings}
\usepackage{xcolor}

\definecolor{codegreen}{rgb}{0,0.6,0}
\definecolor{codegray}{rgb}{0.5,0.5,0.5}
\definecolor{codepurple}{rgb}{0.58,0,0.82}
\definecolor{backcolour}{rgb}{0.95,0.95,0.92}

\lstdefinestyle{mystyle}{
  commentstyle=\color{codegreen},
  keywordstyle=\color{magenta},
  numberstyle=\tiny\color{codegray},
  stringstyle=\color{codepurple},
  basicstyle=\ttfamily\footnotesize,
  breakatwhitespace=false,         
  breaklines=true,                 
  captionpos=b,                    
  keepspaces=true,                 
  numbers=left,                    
  numbersep=5pt,                  
  showspaces=false,                
  showstringspaces=false,
  showtabs=false,                  
  tabsize=2
}

\definecolor{blue}{rgb}{0,0,1}

\newcommand \hl[1]{{#1}}

\begin{document}

%
\newcommand{\frameworkName}{The CARP Mobile Sensing Framework}
\title{\frameworkName -- A Cross-platform, Reactive, Programming Framework and Runtime Environment for Digital Phenotyping}
\renewcommand{\shorttitle}{\frameworkName}

\author{Jakob E.~Bardram}
\email{jakba@dtu.dk}
\orcid{0000-0003-1390-8758}
\affiliation{%
  \institution{Department of Health Technology, Technical University of Denmark}
  \streetaddress{Richard Pedersens Plads}
  \city{Kgs. Lyngby}
  \postcode{DK-2800}
  \country{Denmark}
}


\begin{abstract}
Mobile sensing -- i.e., the ability to unobtrusively collect sensor data from built-in phone sensors -- \hl{has long} been a core research topic in Ubicomp. A number of technological platforms for mobile sensing have been presented over the years and a lot of knowledge on how to facilitate mobile sensing has been accumulated. 
This paper presents the \ac{CAMS} framework, which
is a modern cross-platform (Android / iOS) software architecture providing a reactive and unified programming model that emphasizes extensibility, maintainability, and adaptability.
\hl{Moreover, the \ac{CAMS} framework supports sensing from wearable devices such as an \ac{ECG} monitor, and configuring data transformers. The latter allows to transform collected data to a standardized data format and to implement privacy-preserving data transformations.}
The paper presents the design, architecture, implementation, \hl{and evaluation} of \ac{CAMS}, and shows how the framework has been used in \hl{two} real-world mobile sensing and \ac{mHealth} applications. 
%
We conclude that \ac{CAMS} provides a novel cross-platform application programming framework which has proved mature, stable, scalable, and flexible in the design of digital phenotyping and \ac{mHealth} applications.
\end{abstract}

%
\ccsdesc[500]{Human-centered computing~Ubiquitous and mobile computing}
\ccsdesc[500]{Software and its engineering~Development frameworks and environments}
\ccsdesc[100]{Applied computing~Health informatics}

%
\keywords{mobile sensing, wearable sensing, context-aware computing, mobile health, mHealth, digital phenotyping, sensors, electrocardiography, ECG, eSense}

\maketitle

\section{Introduction}
\label{sec:introduction}

%
%
%

Research in mobile sensing -- i.e.\ the collection of data from sensors in mobile technologies -- has shown that indicators of behavioral, social, psychological, and health status can be derived by collecting continuous and real-world data and applying advanced algorithms to it~\cite{Lane2010}.
%
A significant body of research has been applying mobile sensing to health and wellness applications~\cite{bardram2016personal}, including, for example, the EmotionSense~\cite{lathia2013smartphones}, BeWell~\cite{lane2011bewell}, and StudentLife~\cite{Wang2014} systems that classify physical activity, sleep, and social interaction based on sensor data. Studies in mental health have demonstrated correlations and predictive power between phone-based features on physical activity, mobility, social activity, phone usage, and voice data on the one hand, and mental health symptoms in e.g., depression~\cite{Saeb2016}, bipolar disorder~\cite{grunerbl2015smartphone,Faurholt-Jepsen2016}, and schizophrenia~\cite{barnett2018relapse} on the other.
\hl{In health sciences, mobile and wearable sensing has been defined as central to the `Precision Medicine Initiative'~\cite{collins2015new}; genotypic information is only powerful if phenotypic information is also available~\cite{doi:10.1001/jama.2015.3595}. 
Using wearable and mobile devices, the prospect of testing intervention strategies in a large population becomes appealing. For example, the `MyHeartCounts' study on cardiovascular mobile health has recruited 40,000 smartphone users~\cite{mcconnell2017feasibility}.}
\hl{This} use of everyday mobile and wearable technology for \hl{collecting} behavioral, psychological, and health data has been termed `digital phenotyping'~\cite{Jain2015,Onnela2016,Huckvale2019}, which can be defined as: 
\textit{continuous and unobtrusive measurement and inference of health, behavior, and other parameters from wearable and mobile technology.}

In order to support \hl{easier deployment and configuration} of mobile sensing studies, a number of research projects have aimed at providing more general-purpose support for mobile sensing, including support for configuration of sampling protocols, accessing low-level sensor data, and handling and storing this data. 
Most of this research has focused on providing easy-to-use \textit{platforms} for collection of data from mobile phones and storing this in a cloud-based infrastructure. These platforms typically have the options to configure a sampling protocol or `study', enroll a set of participants, deploy the study onto the participants' mobile phones and automatically collect data in a cloud infrastructure, which can be accessed from a web portal. Contemporary examples of this approach include Purple Robot~\cite{purple}, Sensus~\cite{xiong2016sensus}, the AWARE Framework~\cite{ferreira2015aware}, the Beiwe Research Platform~\cite{torous2016new}, mCerebrum~\cite{Hossain:2017:MMS:3131672.3131694}, RADAR-base~\cite{info:doi/10.2196/11734}, \hl{and LAMP~\cite{Torous2019}} which all are quite elaborate and mature by now. 
These platforms are designed for research and target experimental behavioral researchers as end-users; the goal is to allow researchers to easily configure a study, enroll participants, deploy the study on the participants' phones, and collect the data automatically with as little interaction with participants as possible. 

However, we have found that often there is a need for designing custom-purpose apps for particular domains and patient groups, and to be able to add support for mobile and wearable sensing to such special-purpose apps. Often the motivation for participants to engage in these studies relies on that `there's something in it for them'~\cite{Bardram2017}, which again means that the app should not be designed with the researcher in mind, but the participant. 
Therefore, there is a need for having a mobile and wearable sensing programming framework that allows researchers to easily add data sampling to their own app during design and implementation.

For this purpose we have designed, implemented, \hl{evaluated}, and released the \acf{CAMS} programming framework. This framework is part of the overall \acf{CARP} and has been evolving over six major releases, and has been used in the design and implementation \hl{of two} released \ac{mHealth} apps for mental health and cardiovascular diseases. 
%
%
%
Similar to other platforms, \ac{CAMS} is a state-of-the-art mobile sensing platform with support for collection of a wide range of data types (see Table~\ref{tab:measures} for an overview), and with support for various non-functional features like privacy protection and adaptive sensing. 
%
%
But \ac{CAMS} also has a set of unique design goals and novel technological contributions:

\begin{itemize}

    \item \textbf{Reactive \ac{API}} -- The \ac{CAMS} programming \ac{API} is designed to be simple, yet expressive for the programmer. It supports a modern reactive, stream-based programming model allowing for non-blocking sensing and data processing. This is important in order to avoid \hl{interference of data sampling with the responsiveness of the app}.

    \item \textbf{Unified sensing} -- \ac{CAMS} has a unified sensing concept for both on-board and off-board (e.g. wearable) sensing. This allows for a unified way to set up and use data sampling in the app without the need to consider differences in the underlying \ac{OS} or device.
    
    \item \textbf{Cross-platform} -- \ac{CAMS} supports a unified programming \ac{API} and programming language for both Android and iOS programming. Hence, the same code base and programming \ac{API} is used \hl{to develop apps for} both Android and iOS. 

    \item \textbf{Data Management} -- \ac{CAMS} has extensive support for different data sampling, transformation, storage, and upload methods. A uniform approach to data transformation is adopted, which supports both privacy-enabling transformations as well as transforming data to make it comply with official data standards.
    
    \item \textbf{Extensible} -- Most importantly, \ac{CAMS} is highly extensible in a number of ways: it allows for implementing new data sampling methods (including both phone sensing, external wearable devices, and cloud-based services); it supports creation of new data transformers (both for privacy reasons and data standards), and it allows for creating custom data managers, which can upload data in a specific format to specific servers (or other kinds of data offloading). 
    
    \item \textbf{Maintainable} -- \ac{CAMS} integrates with a publicly available online build and dependency management system which allow for easy access to \ac{CAMS} libraries and for sharing custom libraries amongst a \ac{CAMS} programming community.
    

\end{itemize}

\hl{
The basic scenario \ac{CAMS} seeks to support is to allow programmers to design and implement a custom cross-platform \ac{mHealth} app, which focuses primarily on providing functionality to the patient. To this end, CAMS enables programmers to add mobile sensing capabilities in a flexible and simple manner to application-specific code bases.
This includes configuring collection of health and behavioral data like \ac{ECG}, location, activity, and step count; 
formatting this data according to different health data formats; 
using this data in the app (e.g.~showing it to the user); 
and uploading it to a specific server, using a specific \ac{API} (e.g.~REST), in a custom format. 
}

\section{Evaluation of Related Frameworks}

\hl{The design of \ac{CAMS} followed a user-centered design methodology -- with users being programmers -- in which we first did a thorough assessment of existing frameworks for mobile sensing, including installing the most mature frameworks and using these for system development.}
%
%
We \hl{interviewed} fellow researchers and industrial software engineers, who had been using some of these frameworks and solicited their experience. Our starting point was that we wanted to use an existing framework in our research, and wanted to find the `best fit'. Unfortunately, our research found none of the existing frameworks adequate for our purpose, and this was the reason and motivation for creating \ac{CAMS}. 
This section outlines which mobile sensing platforms and programming frameworks already exist and discusses some of their \hl{strengths} and shortcomings \hl{based on our review}. 



\subsection{Mobile Sensing Research Platforms}

A number of mobile sensing \textit{platforms} exist, i.e.~systems which are built and used for data collection via mobile phones but are not designed to be extended or used as programming frameworks. These systems typically follow the same model: they have a standard mobile phone app (sometime different apps for different data types) which can be configured to collect different types of data, which is uploaded to a cloud-based infrastructure. Often the phone app runs in the background and passively samples sensor data while also asking users to fill in surveys or input other user-generated data, such as subjective stress score. 
The most prominent and mature examples of such platforms include (in chronological order) EmotionSense~\cite{lathia2013smartphones}, AWARE~\cite{ferreira2015aware}, Purple Robot~\cite{purple}, Beiwe~\cite{torous2016new}, Sensus~\cite{xiong2016sensus}, mCerebrum~\cite{Hossain:2017:MMS:3131672.3131694}, RADAR-base~\cite{info:doi/10.2196/11734}, \hl{and LAMP~\cite{Torous2019}}.
%
Most of these systems \hl{only support sampling data from built-in phone sensors (e.g.~accelerometer data, step count, location, etc.), but a few} like RADAR-base and mCerebrum also support data collection from wearable devices. 
\hl{Many} of these systems only support Android, \hl{with the exception of} AWARE, Beiwe, and Sensus, which also \hl{come} with an iOS client app.
Most of the platforms included here are to a large degree \hl{released as open source}\footnote{Some platforms -- like RADAR-base -- keep integration to wearable devices as closed source.} and the more recent ones (AWARE, Sensus, mCerebrum, Beiwe, and RADAR-base) are actively maintained and used in different studies.
However, even though the source code is available and in principle \hl{anyone} can access most of them, \hl{they are challenging to extend and use without a team with a strong technical background~\cite{Torous2019}}.
\hl{In the phone client apps there are a lot of code dependencies which makes them complicated to build,
and there is often no clear separation of the sensing part from \ac{UI} concerns, which makes it hard to reuse the sensing part of the client code.}

\subsection{Mobile Sensing Programming Frameworks}

\hl{In contrast to mobile sensing platforms, mobile sensing \textit{programming frameworks} are designed to be used by software engineers and app developers to include support for mobile and wearable sensing in their own code base.}
In contrast to mobile sensing platforms -- where the end-user is mainly the researcher running a study -- the end-user of a programming framework is the software developer designing and implementing a mobile sensing or \ac{mHealth} app. 

There has been surprisingly little research on providing such `pure' mobile sensing programming frameworks. 
Two of the original examples include Funf~\cite{aharony2011social} and the EmotionSense Libaries~\cite{lathia2013open}, which are Java-based libraries for Android sensing. They are very similar, providing support for configuration of data sampling from on-board mobile phone sensors and \hl{uploading it as files} to a server.
%
We did an interview with a software engineer who had been using Funf to add data sampling to a research app. He was satisfied with Funf in many respects but also found it rather ``\textit{low-level}'' and ``\textit{inflexible}'' in the sense that it only supports very low-level sensor sampling and has no support for extension, i.e.~ `framework hooks' \hl{which can be used to} `hook' into the programming framework and extend it for your own application purpose.
\hl{Moreover, neither} the Funf or the EmotionSense libraries seem to be maintained anymore.

PrivacyStreams~\cite{li2017privacystreams} is a more recent example of a programming framework for mobile sensing, which focuses on providing a modern reactive programming model and privacy protection of data sampling.
In some respects, PrivacyStreams is similar to the design of \ac{CAMS}: 
PrivacyStreams has an expressive, user-friendly, and well-documented programming \ac{API}, provides a modern reactive programming model, and supports sampling of different types of data which can be reused and transformed in a pipeline architecture. The dedicated focus on privacy protection `at the source' is also a core feature of the framework.
However, instead of hooking into a standard stream model (like the ReactiveX\footnote{\url{http://reactivex.io}} model which is also available for Java), PrivacyStreams has designed its own stream model, which adds \hl{extra overhead for} the programmer who has to learn a completely new programming model. Data transformation in PrivacyStreams is also limited to stream operations like mapping, filtering, sorting, selecting, etc., and does not provide any hooks for transforming data into specific data formats, such as health data formats like \ac{OMH}. Finally, PrivacyStreams is not designed to be extensible and has no direct \ac{API} support for adding new sampling modalities, data transformation, or data management and upload functionality\footnote{PrivacyStreams is open source and of course one can check out the code and add all sorts of additional functionality. However, the main argument is that the framework itself does not support extensibility, such as loading and registering new sampling mechanisms, data transformation, and data upload modules.}.
\hl{PrivacyStreams is implemented in Java and only supports Android.}


Some of the mobile sensing platforms claim that their open source code can be used in the development of custom apps -- most notably AWARE and Sensus. We have investigated this as well.
When using AWARE, there is good support for setting up a study and sampling data, and AWARE can be extended with custom `plugins' which can be loaded in the AWARE client app. 
AWARE also comes with an \ac{API} and tutorials for how to use AWARE in a standalone app for both Android and iOS. This allows for data collection which uses a consistent data format across iOS and Android. As such, we found AWARE to be a very stable and mature framework. We did an interview with a programmer who had used AWARE in a custom app and who had found it useful, especially on Android.
He argued, however, that even though the programming \ac{API} is available on both iOS and Android, he found them very different and inconsistent and therefore for all practical purposes they are two distinct \acp{API}. 
Moreover, \hl{he argued that it was difficult to fit AWARE into the commercial technology stack of the company and their custom app since} there is no direct support for data transformation \hl{to their proprietary data formats and uploading data to a different server than AWARE's, which they did not want to install and maintain as part of their setup}. 
Hence, in the end the programmer ended up not using AWARE and instead collected data directly from the \ac{OS} \ac{API}, formatted it according to their own data format, and uploaded it to their own server with its own custom REST \ac{API}.

As for Sensus, we interviewed two researchers who had implemented an app using Sensus. The support for cross-platform (both iOS and Android) data sampling was found very useful, but \hl{they argued that} the Sensus architecture was \hl{difficult for them} to use in app development. Basically, the developers had to \hl{retrieve the Sensus app source code}, disable the Sensus \ac{UI} code and build a new custom \ac{UI} for the new app \hl{-- they labeled this approach as} `hacking' the existing app.
Moreover, \hl{they found that} Sensus is strongly tied to \ac{AWS} in terms of data upload and there was no simple way to upload data to other servers. 

\subsection{Summary}

Most of the available software for mobile sensing is tied to a dedicated client and \hl{back end} infrastructure, which are often designed and build for \hl{specific} research purposes. These platform are great for non-programmers and clinical researchers for engaging in digital phenotyping and many of these express a high degree of maturity in terms of features, maintenance, robustness, documentation, and support. However, these platforms are not designed as programming frameworks to be extended and used in other custom applications by $3^{rd}$ party developers. 
Our literature and code review of the most mature mobile sensing programming frameworks combined with \hl{interviews with programmers with} hands-on experience in using them for application development revealed a set of shortcomings. Except for Sensus, there are no cross-platform frameworks available, and the Sensus programming model turned out to be hard to use.
And even though the AWARE client exists on both Android and iOS, the AWARE programming framework is very inconsistent across Android and iOS. 
\hl{Android-based} frameworks like Funf, the EmotionSense libraries, and PrivacyStreams provide programming \acp{API}, but on a very low level requiring the application programmer to implement a lot of higher-order functionality for data management him/herself.
But most importantly, all of the frameworks had very limited `hooks' for extensibility, which significantly hampered the ability to implement new sensing capabilities, data formatting, transformation, management and upload. Hence, it was very difficult to tailor and use these frameworks for building custom \ac{mHealth} applications for specific diseases and disorders. 



\section{CAMS Application Programming Model}

The main purpose of \ac{CAMS} is to provide a unified and reactive programming \ac{API} for cross-platform sensing. \hl{How} code is written hence plays a core role in the framework. \hl{Therefore,} a natural starting point is to see how \ac{CAMS} is used from an application programmer's \hl{point of view}. 
\ac{CAMS} is implemented in Flutter, which is a cross-platform \hl{software development} toolkit for building natively-compiled mobile applications for iOS and Android. Flutter uses the Dart programming language, which is a modern object-oriented, reactive, non-blocking language.
This section outlines how mobile sensing can be added to a Flutter app.

\subsection{Code Example}

Listing~\ref{code:example} shows how sampling can be added to a Flutter app. 
This basic example illustrates how sampling is configured, initialized, started, and used in four basic steps:
(i) a study is defined (line 4--19); 
(ii) the runtime environment is created and initialized with this study configuration (22--23); 
(iii) the stream of sampling events is consumed and used in the app (26--28);
(iv) data sampling is started (31) and can be adapted at runtime (34).
In the following, we shall dig into the details of how this is achieved by the \ac{CAMS} framework.

\begin{lstlisting}[
language=Java, 
caption={A simple Dart program setting up phone sensing in \acf{CAMS}.}, 
label=code:example]
import 'package:carp_mobile_sensing/carp_mobile_sensing.dart';

void sensing() async {
  Study study = Study('ex-1', 'user@gmail.com',
      name: 'A simple example study',
      dataEndPoint: FileDataEndPoint()
        ..bufferSize = 500 * 1000
        ..zip = true
        ..encrypt = false)
    ..dataFormat = NameSpace.OMH
    ..addTriggerTask(
        ImmediateTrigger(),
        Task('One Common Sensing Task')
          ..measures = SamplingSchema.common().getMeasureList([
            ConnectivitySamplingPackage.BLUETOOTH,
            ConnectivitySamplingPackage.CONNECTIVITY,
            SensorSamplingPackage.ACCELEROMETER,
            SensorSamplingPackage.GYROSCOPE
          ]));

  // create and initialize the sampling runtime with the study
  StudyController controller = StudyController(study);
  await controller.initialize();

  // subscribe to events from the controller
  controller.events.listen((Datum datum) {
    // use the collected data in the app, e.g. show it in the UI
  });
    ...
  // when ready, start sampling
  controller.start();
   ...
  // pause / resume sampling
  controller.pause();
}
\end{lstlisting}

\subsection{Software Domain Model}
\label{sec:domain}

\hl{
The \ac{CAMS} domain \ac{API} consists of three main parts: study, sampling schemes, and data points.
}

\subsubsection{Study}
\label{sec:study}

\begin{figure}[t]
\centering
\includegraphics[width=0.6\textwidth]{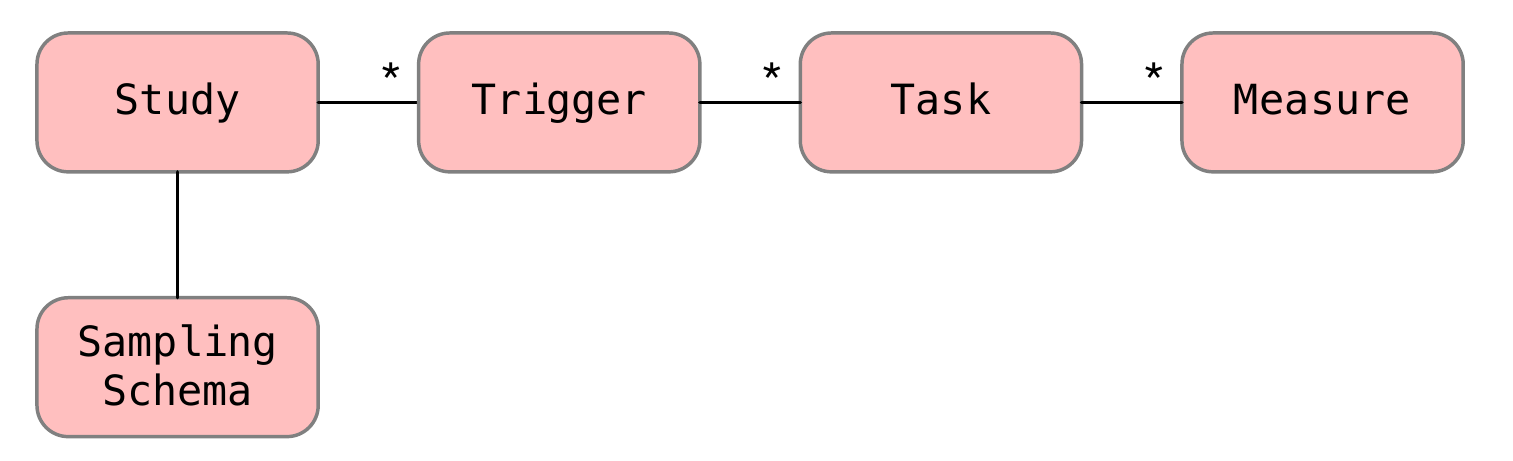}
\caption{The domain model of \hl{\texttt{Study} in} \acf{CAMS}.}
\label{fig:carp_domain}
\end{figure}

\hl{Fig.~\ref{fig:carp_domain} shows the \texttt{Study} part of the \ac{CAMS} domain model, which specifies how} a sampling study is defined and configured. 
A \texttt{Study} consists of a set of \texttt{Trigger}s, which points to a set of \texttt{Task}s, which defines a set of \texttt{Measure}s to be done. 
Compared to other sensing frameworks (which typically just have a list of measures), this model may seem a little over-complicated. However, each type has a specific purpose: 

\begin{itemize}
    \item \texttt{\textbf{Study}} -- The \texttt{Study} holds the entire definition of the study to be done. As shown in Listing~\ref{code:example}:4--9 a \texttt{Study} object defines the \texttt{id}, \texttt{username}, \texttt{name}, and \texttt{DataEndpoint} of the study. The data endpoint \hl{specifies} where to `hand over' the data \hl{to. For example,} to a file or a cloud-based infrastructure. This will be discussed further in section~\ref{sec:data_managers}.

    \item \texttt{\textbf{Trigger}} -- A \texttt{Trigger} defines \textit{when} sampling is done and hence describes the temporal configuration of a study. \ac{CAMS} comes with a set of built-in triggers, including triggers that starts sampling immediately (the \texttt{ImmediateTrigger} used in Listing~\ref{code:example}:12), a trigger than runs periodically with a fixed period (\texttt{PeriodicTrigger}), a trigger that can be scheduled on a specific date and time or \hl{following a recurring pattern} (the \texttt{ScheduledTrigger} and \texttt{RecurrentScheduledTrigger}), and a trigger that starts when a certain sampling event happens, such as when the phone enters a certain geofence (the \texttt{SamplingEventTrigger}).

    \item \texttt{\textbf{Task}} --
    A task is a bundle of measures to be done simultaneously \hl{once initiated by a trigger.}
     For example, a task can specify to sample accelerometer and gyroscope data when triggered during a tremor assessment for Parkinson.
    In Listing~\ref{code:example}:13-19, one task sampling Bluetooth, connectivity status, accelerometer, and gyroscope data is defined.

    \item \texttt{\textbf{Measure}} -- A measure defines \textit{what} to measure, i.e. the specific data stream. A measure is a configuration of the \texttt{type} of data to collect, which during runtime maps to a specific probe that can collect this type of data. Since \ac{CAMS} follows a reactive programming model, all sampled data is collected in \textit{streams} by listening to the underlying sensors. 
    \hl{Measures often need detailed configuration, such as specifying the scanning frequency for Bluetooth.}
    However, this low-level configuration is often irrelevant to the user (programmer) of \ac{CAMS} and can therefore be \hl{abstracted away} using a \texttt{SamplingSchema}. In Listing~\ref{code:example}:14, the \texttt{common} sampling schema is used to get a set of measures with the most `common' configuration. Sampling schemes \hl{are} further described in section~\ref{sec:sampling_schema}.
    
\end{itemize}

%
The \texttt{Trigger} concept is \hl{a powerful construct, as it allows configuring the detailed scheduling of} measures into \hl{separate} \texttt{Task}s.
\hl{For example}, \textit{interval-contingent} and \textit{event-contingent} \acp{EMA}~\cite{wheeler1991self} can be designed by triggering a \texttt{survey} measure using a \texttt{ScheduledTrigger} and \texttt{SamplingEvent Trigger}, respectively.
More advanced scenarios include: `start sampling of Bluetooth and Wifi when entering the home of the user' or \hl{`trigger an assessment of Parkinson's Disease symptoms every day at 8pm'. The Parkinson's Disease assessment task could collect measures like a survey, a tremor test (using \ac{IMU} sensors), \ac{PPG}, \ac{ECG}, and ambient light~\cite{Bloem2019}.}
\hl{A study can be specified in two ways: programmatically using the \ac{CAMS} domain model \ac{API} (as shown in Listing~\ref{code:example}:4--19) or using a JSON file.}

\subsubsection{Sampling Schemes} 
\label{sec:sampling_schema}

In order to hide the complexity of configuration of measures, \ac{CAMS} offers the abstraction of a \texttt{SamplingSchema}, which can be used to set up a study in a much more convenient manner. 
A sampling schema defines a specific configuration of a measure -- for example that the \texttt{bluetooth} measure should scan for proximal bluetooth devices every 10 minutes for 5 seconds (as suggested in the StudentLife study~\cite{Wang2014}).
A \texttt{SamplingSchema} can be created with a default sampling configuration called the \texttt{common} schema.
In Listing~\ref{code:example}:14 the \texttt{SamplingSchema.common()} constructor is used to get a list of measures with the `common' or default\footnote{The keyword \texttt{default} is reserved in Dart 
and hence could not be used.} configuration. 
In addition to the default sampling schemes, programmers can defined custom schemes for specific purposes that can be reused. For example, in the \hl{power consumption} studies of \ac{CAMS} (see section~\ref{sec:technical_evaluation}), two sampling schemes for the two studies were defined, each resembling \hl{the studies of AWARE and mCerebrum as reported in the literature}.
Sampling schemes are defined for each sampling package, which are described in section~\ref{sec:sampling_packages}.

\subsubsection{Data Points} 

A data point holds sampled data and consists of a header specifying the \texttt{DataFormat} and a body, called \texttt{Datum}.
The \texttt{DataFormat} specifies the namespace and the type of the data being sampled. For example, \texttt{carp.bluetooth} specify a data point of type \texttt{bluetooth} in the \texttt{carp} namespace, whereas \texttt{omh.bloodpressure} specify a blood pressure measure in the \acf{OMH} namespace.
The \texttt{Datum} class defines the specific data point value. For example, the \texttt{BluetoothDatum} holds the Bluetooth device name, id, type, power level, and RSSI signal strength.
Similarly, the \texttt{BloodPressureDatum} holds information on the systolic and diastolic blood pressure, and the body position of the measurement\footnote{See \url{https://www.openmhealth.org/documentation/\#/schema-docs/schema-library} for definitions of the \acf{OMH} data schemes.}.

\subsection{Data Transformation}

\hl{In order to support the design of custom \ac{mHealth} apps, the ability to transform data into another format is useful for several reasons}. 
First, transformation may be needed to support a specific data format when uploading to a certain \hl{data storage}. For example, if uploading data to the mCerebrum Cortex \hl{server}, we need to comply to the mCerebrum data schemes. 
Second, transformation is useful when data is to be saved in an standardized format, \hl{such as} the \hl{\acf{OMH}} or \acf{FHIR} data schemes.
Third, transformation may be needed for \textit{privacy} \hl{reasons} and can be used to anonymize data \hl{prior to} upload. 

\begin{figure}[t]
\centering
\includegraphics[width=0.9\textwidth]{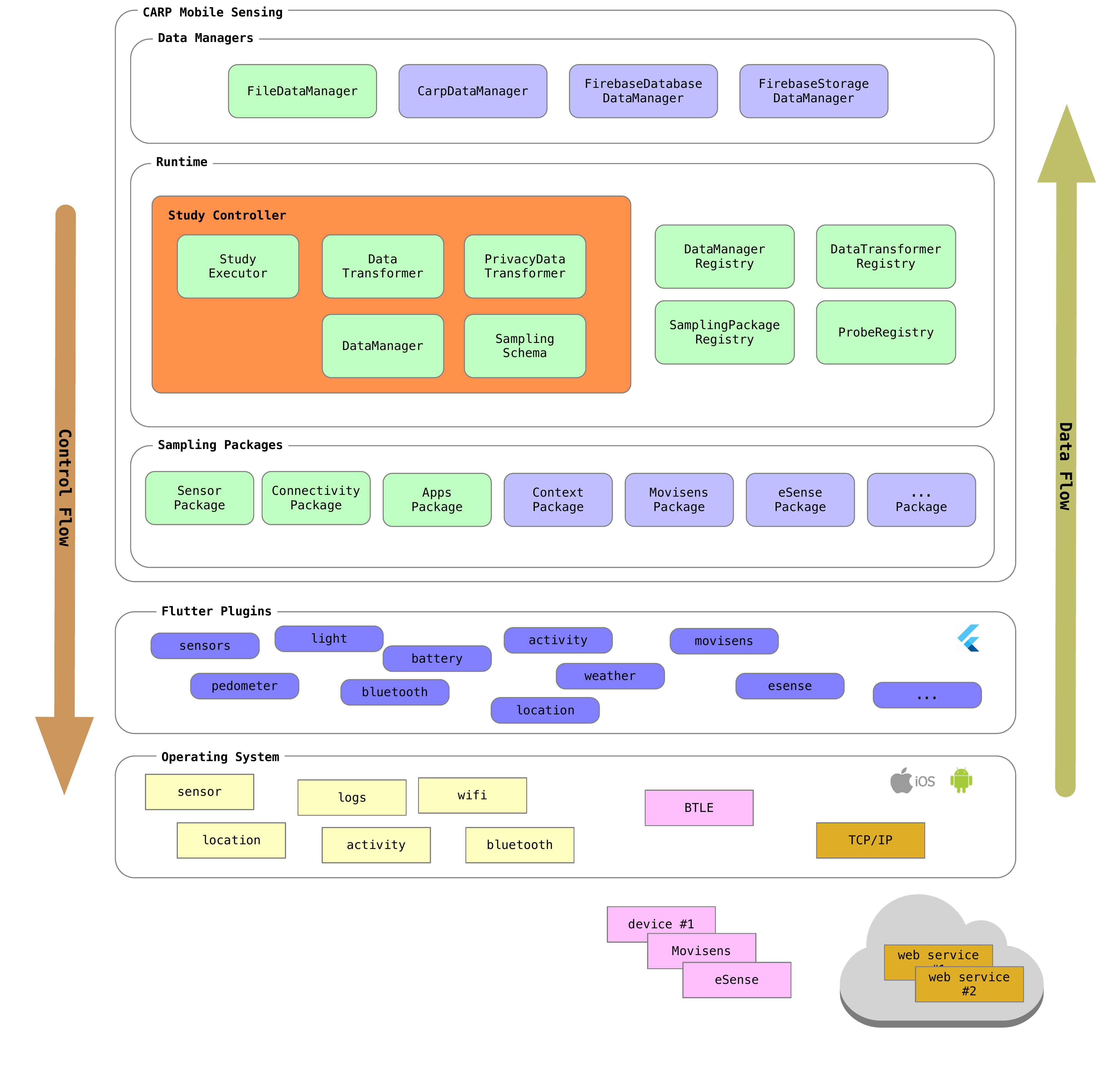}
\caption{The overall runtime architecture and main components of the \acf{CAMS} framework \hl{consist} of three main layers -- runtime, sampling packages, and data managers -- which \hl{build} on top of Flutter plugins and the processes and services in the native \ac{OS}. Sampling is controlled from the study controller and down to the \ac{OS} whereas data flows from \ac{OS} sensors, external wearable devices, and web \hl{services} up towards the data managers.}
\label{fig:carp_architecture}
\end{figure}

\hl{
\ac{CAMS} has a very generic model for creating data transformers, which consists of three main parts.
First, at the lowest level a \texttt{DatumTransformer} function which can transform a \texttt{Datum} object from one data format to another can be defined. 
Second, such datum transformers can be organized into a \texttt{DataTransformer}, which hold a set of \texttt{DatumTransformer} functions for a specific data format. 
Third, data transformers}
can be registered in the \texttt{DataTransformerRegistry} (see Fig.~\ref{fig:carp_architecture}) and used for data transformation at runtime. 
It is beyond the scope of this paper to \hl{describe} how to design, register, and use data \hl{transformation} in \ac{CAMS}, but a tutorial in the online documentation explains how to do this~\cite{carp_documentation}.

\ac{CAMS} comes with three built-in data \hl{transformers}: a CARP, an \ac{OMH}, and a privacy \hl{transformer}, which can be used in the motivating examples above. 
\hl{For example, listing~\ref{code:example}:10 shows how the data format can be specified for a study. In this case, all data points will be transformed into the \ac{OMH} data format by using the built-in \texttt{OMHDataTransformer}.}

\hl{The built-in \texttt{PrivacyDataTransformer} defines a set of functions that can obfuscate data points. If privacy is enabled, this schema is used to obfuscate data points when collected and before any other data transformation or upload is done.}
A common strategy in other mobile sensing frameworks is to support privacy `by default' by scrambling data \hl{at} the source, i.e., when data is being sampled. For example, Funf, Sensus, AWARE, and mCerebrum all \hl{perform} one-way hashing of sensitive data like telephone numbers and email addresses. 
\hl{However, since the original data might be needed, \ac{CAMS} does not enforce any scrambling of data. Instead, more flexibility is provided by considering privacy protection part of data transformation.}


\subsection{Extending the \ac{CAMS} Domain Model}

Since \ac{CAMS} is designed as a highly extensible programming framework, \hl{all \ac{CAMS} domain model classes} can be extended for domain-specific purposes.
\hl{Specifically, the} \texttt{Study}, \texttt{Trigger}, \texttt{Task}, \texttt{Measure}, \texttt{SamplingSchema}, \texttt{Datum}, and \texttt{DataTransformer} classes -- including the built-in implementations of these -- are \hl{extensible}.
Section~\ref{sec:sampling_packages} describes how \hl{to create new sampling packages in \ac{CAMS} by} extending core classes in the \ac{API}.

\section{CAMS Runtime Architecture}

Fig.~\ref{fig:carp_architecture} shows the overall layered software architecture of \ac{CAMS}.
The \ac{CAMS} runtime model \hl{consists} of three main layers: the core runtime, data managers, and the sampling packages, which again \hl{build} on top of a set of Flutter plugins \hl{which} access processes and services in the underlying \ac{OS}, external wearable devices, and cloud-based services.
This runtime architecture is designed to achieve the non-functional software architecture goals of being highly extensible, cross-platform, and maintainable.

\begin{figure}[t]
\centering
\includegraphics[width=0.9\textwidth]{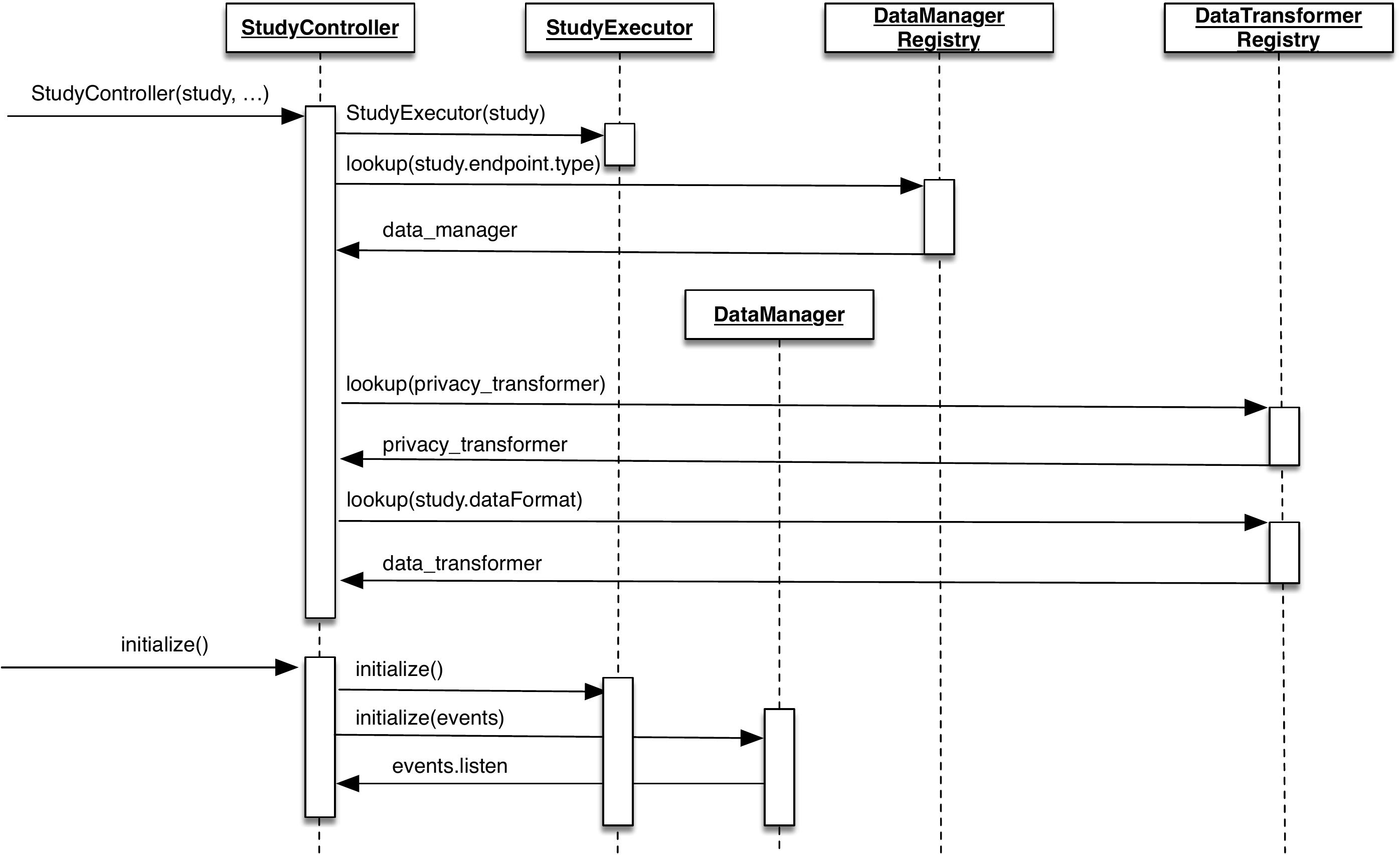}
\caption{The initialization of the \ac{CAMS} runtime.}
\label{fig:cams_init}
\end{figure}

\subsection{Runtime}

As illustrated in Fig.~\ref{fig:carp_architecture}, the \ac{CAMS} runtime make extensive use of registries in which different components can be registered and \hl{retrieved at} runtime. These registries are core to the extensibility of the framework since they allow for adapting and extending how data is acquired, transformed \hl{(including anonymization)}, stored, and uploaded.

As shown in Listing~\ref{code:example}:22, the \texttt{StudyController} is the main runtime component responsible for executing a study, handling sampling packages, transforming data (including applying any privacy \hl{transformer}), and uploading it to a data manager.
Fig.~\ref{fig:cams_init} shows how the \texttt{StudyController} is created and initialized. On creation it (i) creates a \texttt{StudyExecutor}, (ii) looks up a data manager in the \texttt{DataManagerRegistry} based on the type of \hl{data} endpoint, (iii) looks up a privacy \hl{transformer} in the \texttt{DataTransformerRegistry}, and (iv) looks up a \hl{transformation scheme} (also in the \texttt{TransformerSchemaRegistry}) based on the study's data format (e.g.\ \texttt{omh}). 
\hl{After} the study controller is \hl{created}, the study executor and data manager are initialized. Note that the data manager is given the \texttt{events} stream, which \hl{carries} all the sampled data, and the data manager then subscribes (listens) to all data point events.

Once initialized, the \texttt{StudyExecutor} is responsible for executing the data sampling as specified in \hl{the \texttt{Study} domain model.}
\hl{It is beyond the scope of this paper to describe how this is done in detail
\footnote{This is described in the online \ac{CAMS} documentation~\cite{carp_documentation}.}, 
but in short, execution entails looking up appropriate sensing \texttt{Probe}s in the \texttt{ProbeRegistry} and collecting data according to the timing specified by the triggers. Listing~\ref{code:example}:31 shows how sampling is started.}
%
%

\subsection{Adaptive Sampling}

Data sampling can be adapted at runtime. This includes pausing and resuming sampling (Listing~\ref{code:example}:34), as well as adjusting the study configuration including its measures, which in turn will adjust the probes' sampling configuration while the sampling is running.
Adaptive sampling is a very generic property of \ac{CAMS} which is also used for power-aware sampling.
Each \texttt{SamplingPackage} defines \hl{a set of default} sampling schemes \hl{that specify how sampling should be done at different battery levels: \texttt{normal}, \texttt{light}, \texttt{minimum}}, and \texttt{none}.
At runtime, these sampling schemes are used for adjusting sampling; when the battery level decreases / increases to certain thresholds, the runtime environment will adapt the sampling by using an appropriate sampling scheme (battery level in parenthesis): \texttt{normal} (100\%--50\%), \texttt{light} (50\%--30\%), \texttt{minimum} (30\%--10\%), and \texttt{none} (10\%--0\%).

\subsection{Sampling Packages}
\label{sec:sampling_packages}

A core mechanism for enabling the high extensibility and robustness of \ac{CAMS} is the sampling package model and \ac{API}.
\hl{
A sampling package is responsible for specifying what data it can collect, the format of this data, and how data is acquired. A sampling package typically handles the collection of a set of related measure types. For example, the connectivity sampling package collects data on connectivity status, wifi, and bluetooth. 
}
%
%
New sensing capabilities can be added to \ac{CAMS} by creating a new \texttt{SamplingPackage} \hl{implementing} the following classes:

\begin{itemize}
    \item \texttt{SamplingPackage} -- the overall specification of \hl{the sampling} package, \hl{specifying which measures it can collect}, factory methods to get the default \texttt{SamplingSchema}s, \hl{a list of datum transformation functions, including} default privacy-preserving functions.
    \item \texttt{Measure} -- specifies what to collect (i.e., the measure type) and how (i.e., the configuration of the probe).
    \item \texttt{Datum} -- specifies the format of the collected data.
    \item \texttt{Probe} -- \hl{the concrete implementation to collect data} from the underlying \ac{OS} or external services using one or more Flutter plugins.
\end{itemize}{}

It is beyond the scope of this paper to \hl{describe} how these classes are implemented but an online tutorial explains this and provides an example~\cite{carp_documentation}.

In Fig.~\ref{fig:carp_architecture}, the built-in sampling packages are shown in green, whereas the external packages are shown in purple.
External packages are only loaded \hl{if they are used in the} app.
This modular design of \ac{CAMS} \hl{has} a number of benefits, including decreasing app size, reducing \hl{dependencies} to only those packages needed, and \hl{not having to ask the user for permission to access data sources that are unused}.
%
%
The sampling package concept also provides a strong modularization model for adding \textit{external} sampling to an app. For example, real-time \ac{ECG} data collection from the Movisens EcgMove4 device is implemented as a separate sampling \hl{package}, which encapsulates the low-level details of handling this device and its data formats.
%
%
Similarly, support for user surveys has been implemented as a sampling package.

Sampling packages are Dart libraries, which can be released as Flutter packages on Dart \texttt{pub.dev}\footnote{\url{https://pub.dev}}. This means that they can be downloaded and added to a Flutter app as needed when the app is built. It also means -- in contrast to most other sensing frameworks which has all the probes built-in -- that an app developer only needs to download and add sampling packages which are needed for his/her specific app. Hence, if context information is not needed for an app, the context package is not linked and used. 
This also means that application programmers can share \ac{CAMS} sampling packages with each other via the official Dart code sharing infrastructure, which includes continuous quality assessment of the code.
Table~\ref{tab:measures} \hl{in Appendix~\ref{app:measures} shows a list of currently available sampling packages and their measures}.

%

\subsection{Data Managers}
\label{sec:data_managers}


As shown in Fig.~\ref{fig:cams_init}, \hl{during the creation of the study controller, a \texttt{DataManager} is retrieved from the \texttt{DataManager Registry} based on the specified data endpoint type, which again is specified as part of the study \hl{(see Listing~\ref{code:example}:6)}. This data manager then subscribes to the \texttt{events} stream and is responsible for handling incoming data.}
As shown in Fig.~\ref{fig:carp_architecture}, \ac{CAMS} currently supports a file data manager and data managers for cloud storage on Firebase and the \ac{CARP} backend.
%
Note that the three cloud-based managers (shown in purple) are available as external Flutter plugins, which means that they are only downloaded and linked, if used. Hence, an app would typically use only one of these.

\ac{CAMS} supports the creation of new data managers -- or extending the existing ones -- for special-purpose data management. This is done by implementing two base classes: \hl{\texttt{DataEndPoint} -- specifies how a data endpoint is to be configured and \texttt{DataManager} -- implements the details of how to store or upload data from the \texttt{events} stream.}
Moreover, if such data managers are release at Dart \texttt{pub.dev}, they would be available for other programmers using \ac{CAMS}.
\hl{Again, it is beyond the scope of this paper to describe how these classes are implemented, but a tutorial in the online \ac{CAMS} documentation goes into more details and provides an extensive example~\cite{carp_documentation}.}

\subsection{Implementation and Availability}


\ac{CAMS} is implemented in Flutter v1.7 using Dart v2.4. 
Flutter is a cross-platform toolkit for building natively-compiled applications for mobile, web, and desktop from a single codebase~\cite{flutter}.
Dart is a modern object-oriented, reactive programming language optimized for non-blocking \ac{UI} programming with a mature and complete async-await event-driven code style, paired
with isolate-based concurrency model.
The implementation of \ac{CAMS} particularly exploits three core aspects of Dart: 
(i) the asynchronous non-blocking programming style using the \texttt{Future} construct, 
(ii) the event-driven reactive stream model using the \texttt{Stream} \ac{API}, 
and (iii) access to native \ac{OS} processes via the \texttt{PlatformChannel} \ac{API}.

Note that the Flutter plugins shown in Fig.~\ref{fig:carp_architecture} are not part of \ac{CAMS}, but are $3^{rd}$ party plugins available from the Dart packages repository. The \ac{CARP} team contributes to this library of Flutter plugins and has released several. However, they are designed to be general purpose and are not specific to \ac{CAMS} -- \ac{CAMS}-specific use of these plugins happens in the \texttt{Probe} implementation in the different sampling packages.


\ac{CAMS} has been designed, implemented, and tested over the course of six major releases and the core framework and all of its associated plugins, sampling packages, and data backends have been released on the Flutter \texttt{pub.dev} software package repository. The framework has been downloaded and used a number of times.
\hl{We are getting issue reports and feature request from outside our lab, so others are seemingly using it.}
Appendix~\ref{app:online_resources} provides an overview of all the online \ac{CAMS} resources, including \ac{API} documentation and online tutorials.

\section{Evaluation}

\subsection{Technical Evaluation and Benchmarking}
\label{sec:technical_evaluation}

In order to evaluate and benchmark \ac{CAMS}, a set of technical evaluation studies were performed in which the performance of \ac{CAMS} was compared to the AWARE framework~\cite{ferreira2015aware} and the mCerebrum platform~\cite{Hossain:2017:MMS:3131672.3131694}.
All studies used the latest released version (\hl{November} 2019) of the Android client apps of AWARE (v4.0.815), mCerebrum (v2.0.14), and \ac{CAMS} (v0.6.\hl{2}). 
The main purpose of these technical studies was to investigate whether the cross-platform runtime environment of Flutter comes with any significant performance penalties compared to native Android sensing frameworks.


\subsubsection{Memory}

On Android, the size of the \ac{APK} files for the AWARE, mCerebrum, and \ac{CAMS} clients are 6.1 MB, 34.1 MB\footnote{mCerebrum consists of several apps which are designed for different purposes and installed as needed. The minimum setup for mobile sensing are three apps: the mCerebrum app (17.8 MB), the PhoneSensor app (7.6 MB) and the DataKit app (8.7 MB) -- in total 34.1 MB}, and 38.8 MB respectively. 
In comparison, the size of the `Hello World' app provided by Flutter is 35.1 MB; it thus seems cross-platform support in Dart/Flutter comes at the cost of an initial payload of 35 MB. Hence, the additional size of adding \ac{CAMS} sensing capabilities to an app is in the order of 4 MB.
In order to investigate working memory usage, the \ac{CAMS} client app was tested using the Android Profiler~\cite{android_profiler} over the course of two hours. This revealed a stable memory footprint in the range of 45--49 MB while running in the background.
%
\hl{
In comparison, the runtime memory consumption of mCerebrum is 80 MB and 50 MB for AWARE.
Hence, there seems to be no memory penalty in using Flutter and hence \ac{CAMS}.
}

\begin{figure}[t]
\centering
\includegraphics[width=0.8\textwidth]{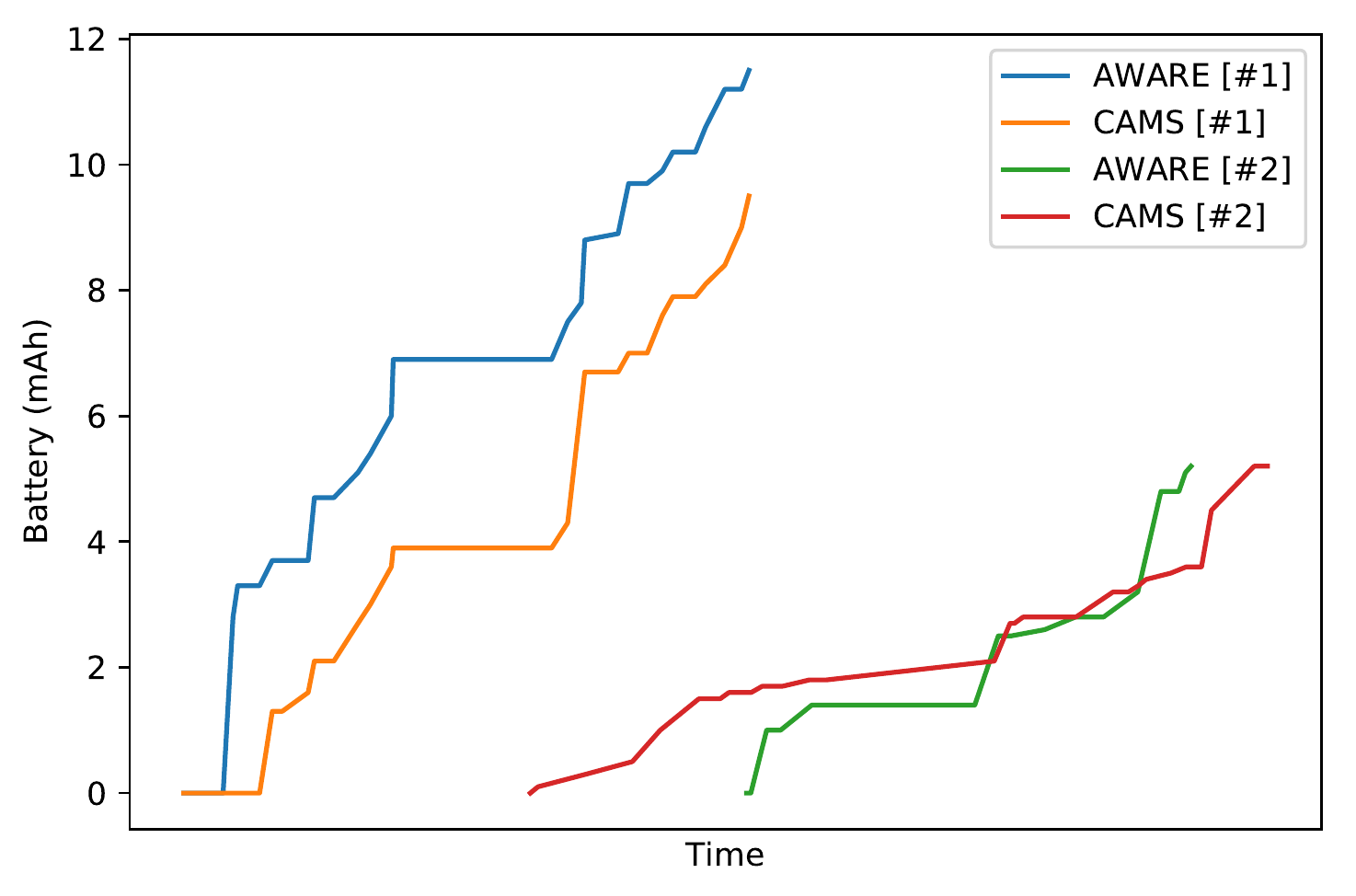}
\caption{Comparing battery consumption of AWARE and \ac{CAMS} during two tests: 
\hl{Test \#1 --  AWARE and \ac{CAMS} running concurrently for 24 hours. 
Test \#2 -- AWARE and \ac{CAMS} running independently for 24 and 36 hours respectively.}}
\label{fig:battery}
\end{figure}

\begin{figure}[t]
\centering
\includegraphics[width=0.8\textwidth]{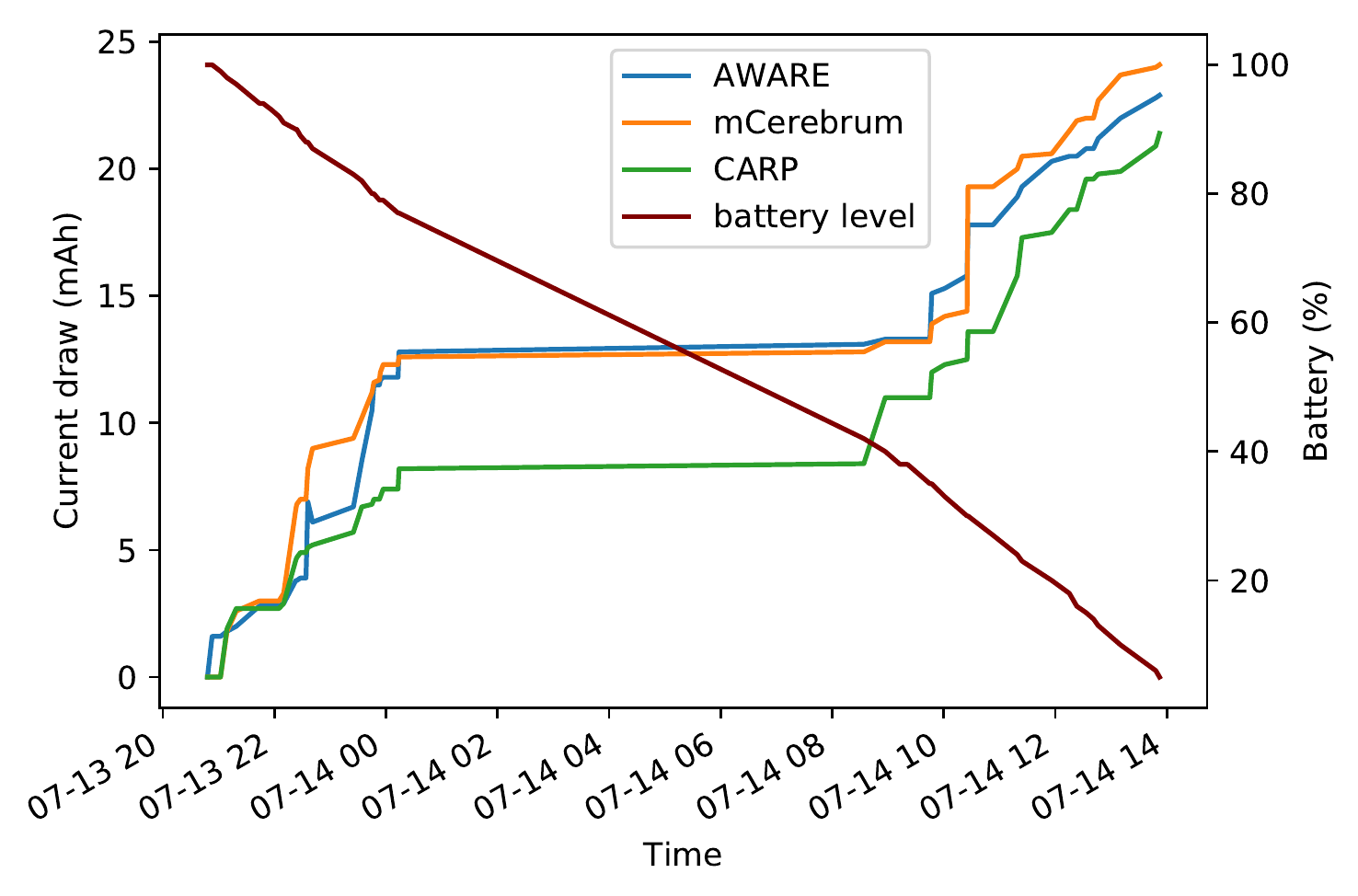}
\caption{Battery consumption during the high-frequency test comparing AWARE, mCerebrum and \ac{CAMS}. 17-hours runtime.}
\label{fig:battery_3_apps}
\end{figure}

\subsubsection{Battery}

Power consumption and battery life are major concerns in mobile sensing~\cite{Lane2010,Hossain:2017:MMS:3131672.3131694}. Therefore, \ac{CAMS} was subjected to \hl{two} power consumption tests: \hl{one mimicking a typical sensing scenario, and another focusing on high-frequency data collection.}
Following the evaluation strategy in \citeauthor{Hossain:2017:MMS:3131672.3131694}~\cite{Hossain:2017:MMS:3131672.3131694}, we compared \ac{CAMS} to AWARE and mCerebrum.
However, in contrast to the power consumption tests done in the \citeauthor{Hossain:2017:MMS:3131672.3131694}~\cite{Hossain:2017:MMS:3131672.3131694} study which only lasted one minute, our studies ran for up to \hl{36} hours in real-world sensing scenarios.
%
\hl{All tests were} done on Android, using a Samsung Galaxy A3, Android OS level 8.0. Power consumption statistics were collected using the AccuBattery Pro app~\cite{accubattery}, which is based on the research of~\citeauthor{choi2002factors}~\cite{choi2002factors}. 
\hl{Power-aware sampling adaptation was disabled in \ac{CAMS}.}
Data was stored locally on the phone using the default storage mechanism of the apps (AWARE uses a database and mCerebrum and \ac{CAMS} use files).

The purpose of the first study was to \hl{compare AWARE to \ac{CAMS} by sampling data from} as many sensors/probes as possible over an extensive period of time.
The common set of data types supported by both apps was identified and each app was configured to collect this data at the same sampling frequency. 
The following sensors were used (sampling period in parentheses, when applicable):
\texttt{accelerometer} (200ms),
\texttt{gyroscope} (200ms),
\texttt{light} (200ms)
\texttt{applications},
\texttt{battery},
\texttt{bluetooth} (60s),
\texttt{communication},
\texttt{location},
\texttt{screen},
\texttt{wifi} (60s),
\texttt{activity} (60s), and
\texttt{weather} (60m).
%
\hl{During a first test,} the two client apps were running concurrently \hl{for 24 hours} in order to be subject to identical sensing scenarios.
\hl{During a second test, the two apps ran independently during two different periods (24 hours for AWARE and 36 hours for \ac{CAMS}).}
Activities during the \hl{two} tests included a mixture of office work, walking, driving, and sleeping. 
%
Fig.~\ref{fig:battery} shows the result of \hl{these two tests}. During the first test (\#1), power consumption of the two apps is very similar, while AWARE seems to consume slightly more power initially. 
\hl{During the second test (\#2) where data collection ran independently, we found no significant difference in power consumption.}
\hl{This study shows} that the energy drain from both \hl{frameworks} is small (below 12 mAh). 

The second study was inspired by the test done by \citeauthor{Hossain:2017:MMS:3131672.3131694}~\cite{Hossain:2017:MMS:3131672.3131694} focusing on high-frequency data. Even though \ac{CAMS} is not designed specifically for high-frequency sampling (like mCerebrum is), it is still of interest to see how it performs under high load. Therefore, a test scenario \hl{similar to the test done by \citeauthor{Hossain:2017:MMS:3131672.3131694}} was created in which the following data were collected at the same frequency:
\texttt{accelerometer} (20ms/50Hz),
\texttt{gyroscope} (20ms/50Hz), and
\texttt{light} (20ms/50Hz).
%
All three apps (AWARE, mCerebrum, and \ac{CAMS}) ran concurrently in order to collect data under identical circumstances. Again, in contrast to the mCerebrum study which only lasted 10 minutes, this study ran for 17 hours involving sensing during a heterogeneous set of real-wold activities as in the first study.
Fig.~\ref{fig:battery_3_apps} shows the result of the study comparing the power consumption of the three apps over the 17 hours study. We observe that the power consumption profile is identical for the three apps, while AWARE and mCerebrum seem to be consuming slightly more power. All three apps was restarted during this long-term test, which can be seen in the graph: AWARE was restarted around midnight on the 13th, \ac{CAMS} at 8 in the morning on the 14th, and mCerebrum at 10 on the 14th. These restarts result in a significant peak in power consumption.

Based on these two studies we can conclude that the cross-platform \ac{CAMS} framework performs equivalently to native Android sensing platforms (AWARE and mCerebrum) and support for cross-platform sensing does not seem to come with a penalty in terms of increased power consumption.

\begin{figure}[t]
\centering
\includegraphics[width=0.75\textwidth]{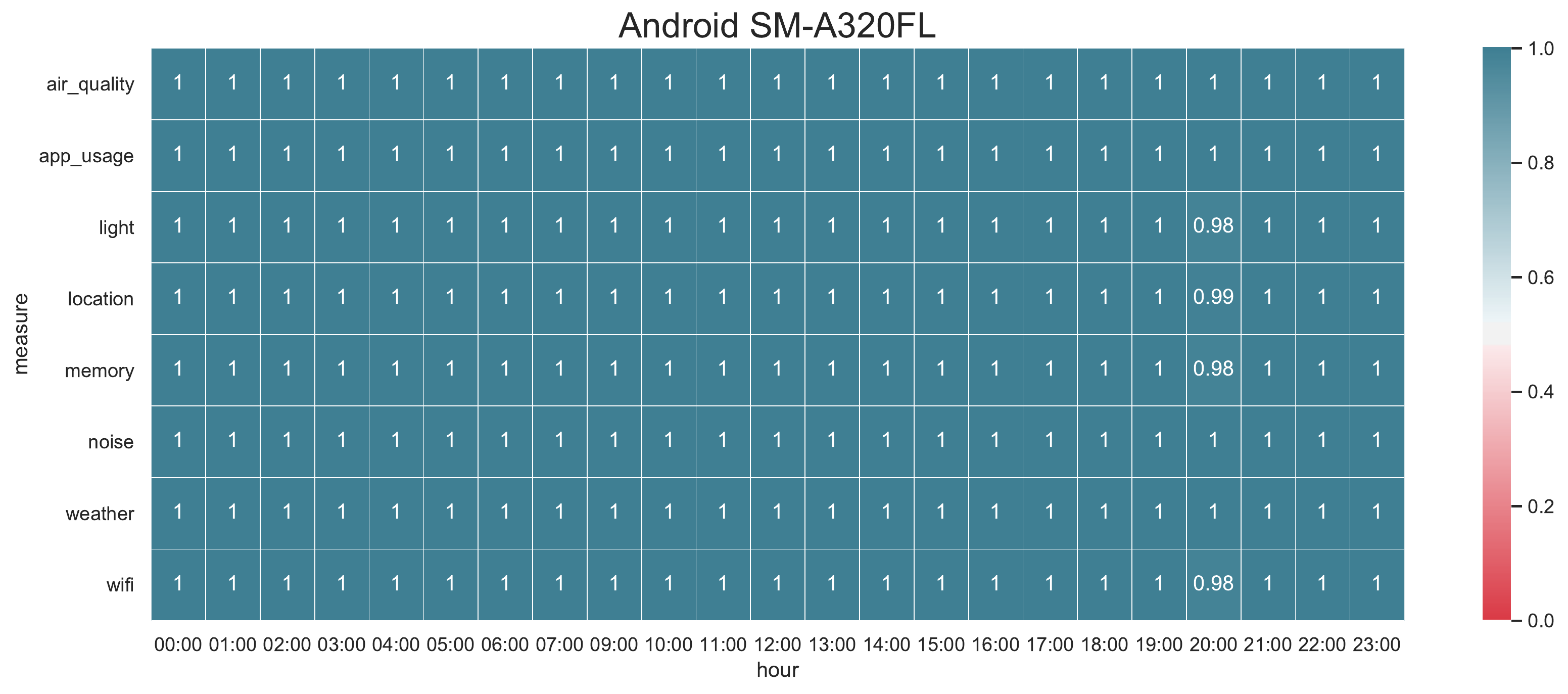}
\includegraphics[width=0.75\textwidth]{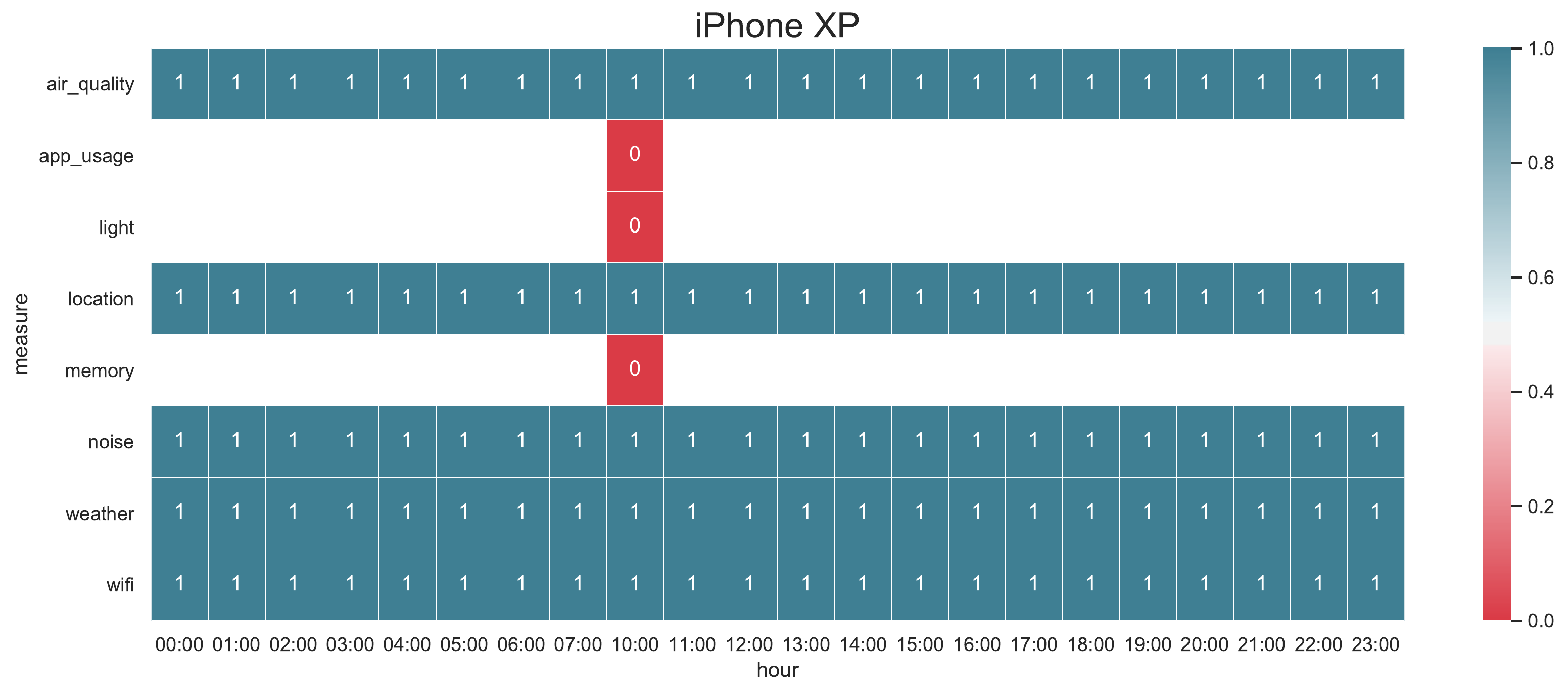}
\caption{\hl{Sampling coverage. Top: Android Samsung A3, sampling period 09:00--08:00. Bottom: iPhone XP, sampling period 10:00--08:00. Numbers in cells indicate sampling coverage; $1$ indicates 100\% coverage, i.e.~that all expected data points were collected. Note that the \texttt{app\_usage}, \texttt{light}, and \texttt{memory} measures are not available on iOS (see Table~\ref{tab:measures}) and hence not collected on the iPhone.}}
\label{fig:coverage}
\end{figure}

\subsubsection{\hl{Sampling Coverage}}

\hl{
A third technical parameter to address is sampling coverage, i.e., to what degree does \ac{CAMS} collect data points compared to what is expected. For example, if the sampling frequency for location is set to 30 s, 120 location data points are expected per hour.
Sampling coverage can only be calculated if ground truth is know, i.e., if we know the expected number of data points. Ground truth can be established if sampling is periodic (set to a fixed frequency) -- such as sampling light at a certain frequency -- but cannot be established if sampling is event driven -- such as activity recognition.
%
}

\hl{Fig.~\ref{fig:coverage} shows the result of a 24-hour coverage test executed on the Android Samsung A3 (top) and an iPhone XP (bottom).
The test includes most of the periodic measures in \ac{CAMS}, for which ground truth can be established via a known sampling frequency: 
\texttt{location} ($0.5$s), 
\texttt{wifi} (1s), 
\texttt{light} (1s), 
\texttt{noise} (1s), 
\texttt{app\_usage} (5s), 
\texttt{memory} (1s), 
\texttt{weather} (10s), and 
\texttt{air\_quality} (10s) (see Table~\ref{tab:measures}).  
%
Coverage is calculated per measure on an hourly basis as the ratio of collected data points compared to the expected number.
%
Overall, the results in Fig.~\ref{fig:coverage} reveal that sampling coverage on both Android and iOS is close to $100\%$ for all measures supported by the respective \acp{OS}. Hence, we can conclude that \ac{CAMS} provides full coverage in data sampling using an identical study configuration across both platforms.
}

\subsection{Application Studies}
\label{sec:cams_app}

\ac{CAMS} has been used in the development of \hl{two real-world \ac{mHealth} apps.}
\hl{These apps have been developed by five experienced developers from the same lab as the authors of \ac{CAMS}, but they are not co-authors of \ac{CAMS} or this paper. 
%
%
Table~\ref{tab:participants} shows an overview and programming background of the programmers. 
All were involved in implementing mobile sensing support in one of the two apps, and one programmer added mobile sensing support for a third app (not discussed in this paper). As part of the the user-centered design of \ac{CAMS}, they were asked to take notes while using \ac{CAMS}, especially regarding the usability and usefulness (pros and cons) of the \ac{API} and online documentation.
Table~\ref{tab:benchmark} shows overall statistics on the two apps, including the number of data points collected and uploaded via \ac{CAMS}. This illustrates that \ac{CAMS} was used in non-trivial apps for collecting a non-trivial amount of mobile sensing data, while also illustrating two very different ways of using \ac{CAMS}.
}

\subsubsection{MUBS}

MUBS is an app supporting \acf{BA}~\cite{rohani2020mubs}, which is a therapy form that emphasizes planning of small, achievable activities as part of treatment for depression. 
MUBS was released in the winter of 2019 and is available on both the Google and Apple app stores. 
At the time of writing, \hl{174} users have registered for MUBS, and ca.~\hl{25 users use it at least once on a weekly basis, with a peak use during a clinical study}.
There are \hl{56}\% Android and \hl{44}\% iOS users, illustrating the cross-platform nature of \ac{CAMS}. 
The majority of users were from the app's home country (\hl{45}\%) \hl{and the \ac{US} (38\%)}.
%
According to the logs, \hl{none of the MUBS users encountered any crashes}, illustrating the stability of the app and hence \hl{also} \ac{CAMS}. 
%

\begin{figure}[t]
\centering
\frame{\includegraphics[width=0.3\textwidth]{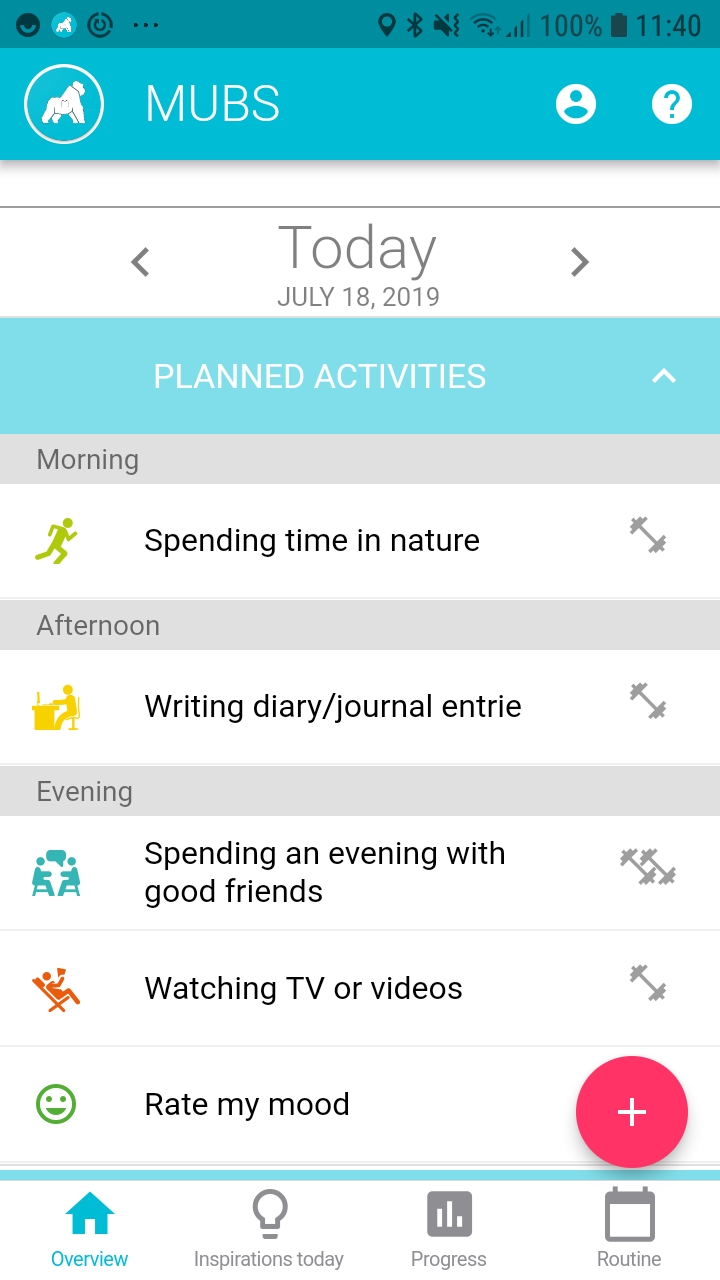}}
\quad
\includegraphics[width=0.5\textwidth]{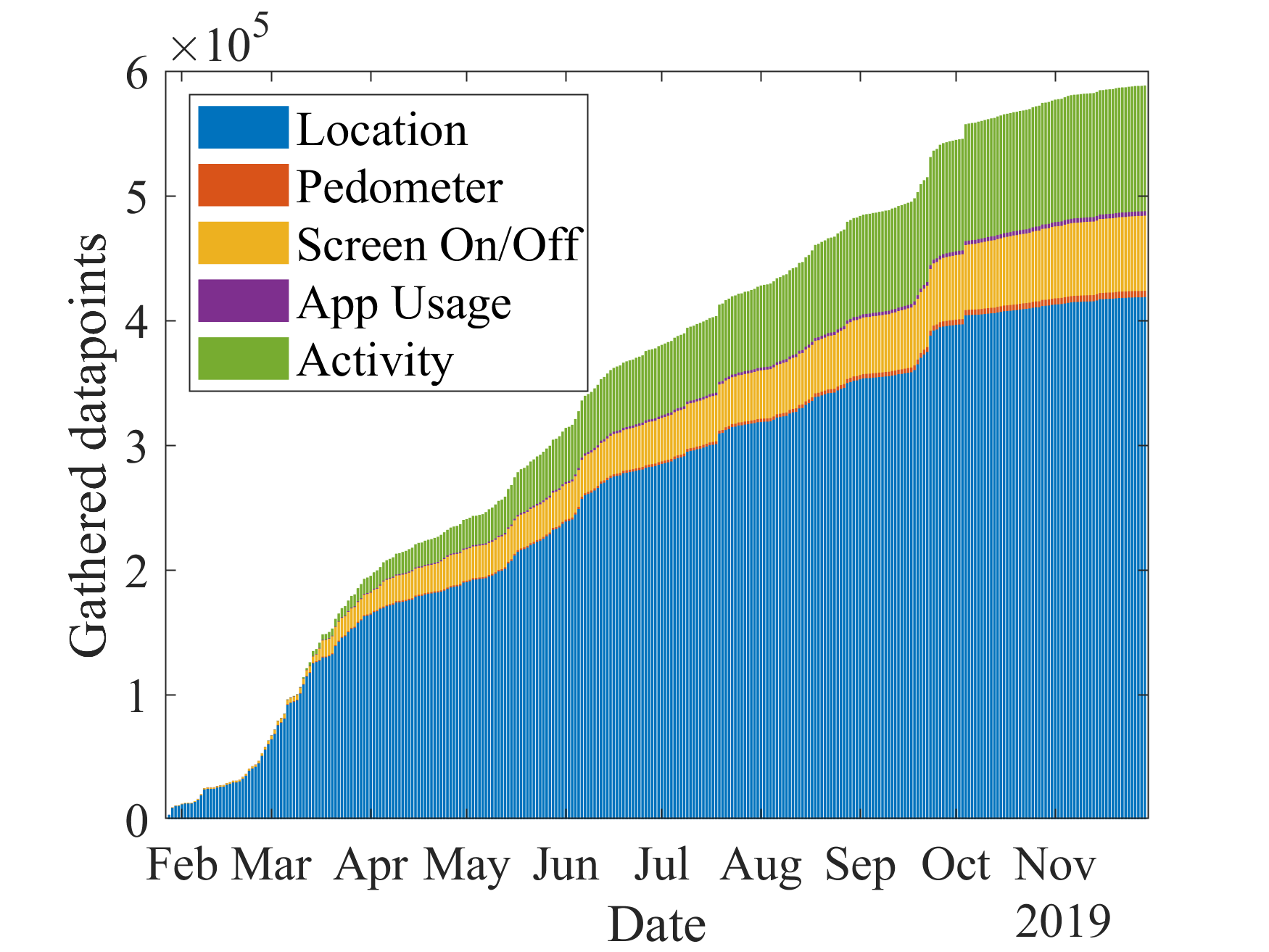}
\caption{\hl{Left:} the user interface of the MUBS app for \acf{BA}. Right: \hl{the collection of data points collected via \ac{CAMS} in the MUBS app since its release in February 2019.}
}
\label{fig:carp_mubs_data}
\end{figure}

\hl{The software architecture of MUBS follows the `InheritedWidget' architecture~\cite{flutter_architectures}, and sensing is added to the app via a sensing utility class.}
MUBS uses \ac{CAMS} to collect data on location, steps, weather, activity, screen, and app usage. 
Once sampling is started it runs continuously. 
All data is uploaded to a Firebase cloud database using the \texttt{FirebaseDatabaseDataManager} plugin. An activity recommendation algorithm is implemented as a Firebase function, which utilize the data being uploaded in real-time. 
\hl{
Fig.~\ref{fig:carp_mubs_data} (right) shows the collection of data points collected via \ac{CAMS} over time. Since a strong relationship between location and depression has been shown~\cite{CanMus15_trajectories,info:doi/10.2196/mhealth.9691}, MUBS collects location at similar high sampling frequency (every 30 secs.). As shown in Table~\ref{tab:benchmark}, MUBS has now been running for $305$ days (Jan.~27--Nov.~28) and close to $600,000$ data points have been collected and uploaded to Firebase. 
The large standard deviations (SD) in terms of data points per day and user reflects that MUBS was used more extensively during clinical trials.
}

\subsubsection{\hl{mCardia}}

\begin{figure}[t]
\centering
\includegraphics[width=0.8\textwidth]{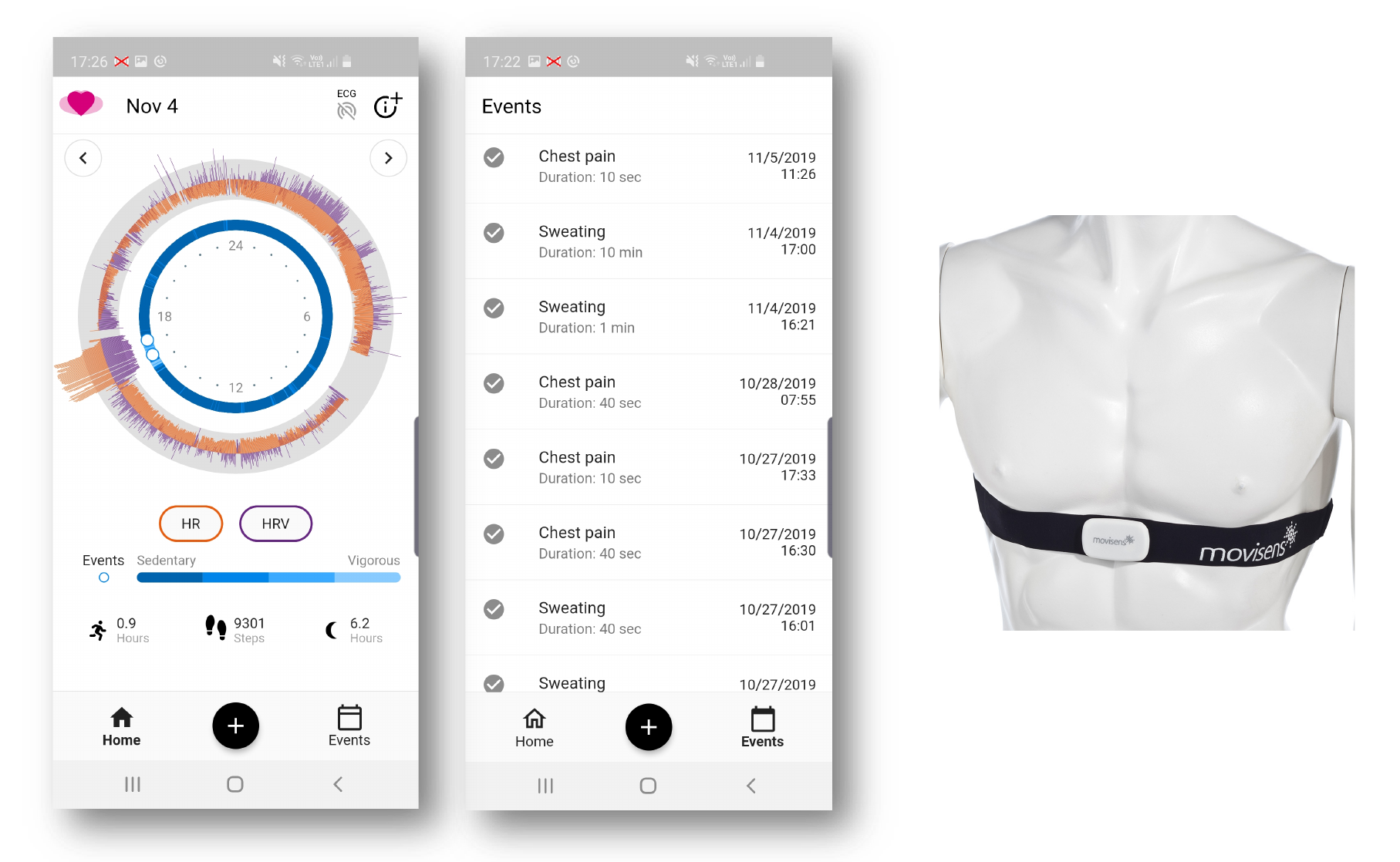}
\caption{The \hl{mCardia} app for \acf{CVD} monitoring -- (i) the homepage of the app showing an overview of the data \hl{being collected via \ac{CAMS} (\ac{HR}, \ac{HRV}, metabolic level, physical activity, and sleep) -- (ii) the list of events registered via tapping the device and sampled via the \ac{CAMS} Movisens sampling package -- (iii)} the Movisens EcgMove4 device for \ac{ECG} monitoring.}
\label{fig:heartwave}
\end{figure}

\hl{mCardia} is an app for monitoring \acp{CVD}. 
\hl{Compared to MUBS, mCardia utilizes \ac{CAMS} beyond just sensing and is a proof of concept of several of its core features, including extensibility.
First, using the prepackaged \ac{CAMS} phone sampling packages, location, activity, steps, noise, \hl{air quality}, and general phone usage is collected.
Second, collection of \textit{high-frequency} data from a wearable \ac{ECG} device is implemented by \textit{extending} \ac{CAMS} with a sampling package for collecting data from the Movisens EcgMove4\footnote{\url{https://www.movisens.com/en/products/ecg-sensor/}} device.
This device collects \acf{HR}, \acf{HRV}, metabolic activity level, steps, and tap markers.
Support for the EcgMove4 device \hl{is implemented via} two Flutter plugins: a generic \texttt{movisens\_flutter} plugin which can access data from the device over \ac{BTLE} using the Android \ac{API} provided by Movisens, and a \texttt{MovisensSamplingPackage} which uses the \texttt{movisens\_flutter} plugin to collect data via \ac{CAMS}. 
Third, \textit{user surveys} about heart problem events such as feeling dizzy or having chest pains are collected. An event can registered by `double tapping' the Movisens device and later the details can be filled in on the phone. 
%
Fourth, in order to collect data in a standardized way for later reuse and cross-validation, data collected in mCardia is \textit{formatted} according to the \ac{OMH} data schemes (where they exists). This is done simply by specifying the study's data format as the \ac{OMH} namespace, as shown in Listing~\ref{code:example}:10.
All data is uploaded to the \ac{CARP} cloud using the \texttt{CarpDataManager}. 
Finally, the software architecture of mCardia follows the \ac{BLoC} pattern~\cite{flutter_architectures}, in which sensing is handled in a centralized sensing \ac{BLoC} component that separates mobile sensing from the rest of the app. This allows the app to \textit{access the data in real-time} using the \texttt{events} stream. This is used to display data in real-time in the user interface, as shown in Fig.~\ref{fig:heartwave}, which shows minute-for-minute \ac{HR}, \ac{HRV}, tap markers, metabolic activity level, active time, steps, and sleep.
}

Currently, Movisens only provides an Android \ac{API} and \hl{mCardia} is hence mainly useful for Android users. However, if Movisens releases an \ac{API} for iOS, support for iOS users would \hl{also be available without any changes to the mCardia implementation}. Moreover, the plugable architecture of \ac{CAMS} allows for supporting other \hl{\ac{HR}} or \ac{ECG} devices by implementing appropriate sampling packages, which can be directly linked and used in the app.

At the time of writing, initial pilot tests of \hl{mCardia} are ongoing. Cardiovascular disease monitoring typically runs over 24--48 hours, and seldom more than 4 days. \hl{As shown in Table~\ref{tab:benchmark},} the app collects high-frequency data from the EcgMove4 device and phone -- up to 15,000 data points per user every 24 hours. Our tests show that the app runs stable during these conditions. Different strategies for increasing robustness and scalability \hl{of \ac{CAMS} have been implemented based on input from the programmers of mCardia}, including buffering of data on the phone prior to upload.

\begin{table}[t]
\caption{\hl{Programmers using} \ac{CAMS}. PE: Programming Experience -- MPE: Mobile Programming Experience -- FPE: Flutter Programming Experience -- Apps: Number of mobile app releases.}
\label{tab:participants}
\small
\centering
\begin{tabular}{l l c c c p{40mm} c}
\toprule
\textbf{ID}  & \textbf{PE} & \textbf{MPE} & \textbf{FPE} & \textbf{Education} & \textbf{Mobile Tech.} & \textbf{Apps}\\
\midrule
P1  &   2-5 &	0-2 &	0-2 &	Master	    &   Xamarin	& 1 \\                          
P2  &   2-5 &	0-2 &	0-2 &	Master	    &   Flutter	& 3 \\                          
P3  &   2-5 &	0-2 &	0-2 &	Bachelor    &	Native iOS, Flutter &	1 \\            
P4  &   2-5 &	0-2 &	0-2 &	PhD	        &   Native Android, React Native &	0 \\    
P5  &   2-5 &	2-5 &	0-2 &	Bachelor    &   Native Android, Native iOS   &	4 \\    
\bottomrule
\end{tabular}
\end{table}

\subsection{\hl{Programmer Feedback}}

\hl{The feedback from programmers helped identify strengths and suggestions for improvement for the \ac{CAMS} \ac{API} and documentation.}
%
%
\hl{One common feature highlighted as important by all programmers was the support for building consistent} cross-platform mobile sensing apps. 
\begin{quote}
\emph{
I'm an iOS guy who is turning to Flutter, and the ability to build [data collection] once is very good. It makes life a lot easier that data is collected in the same way and has the same data format.}
[P3]
\end{quote}
%
%
The way that \ac{CAMS} abstracts away the sampling in one unified and simple \ac{API} is \hl{based on continuous input from the programmers and the design goal was to} let programmers \hl{configure} sampling in a standardized manner across iOS and Android. 
It was highlighted that the design of \ac{CAMS} and its \ac{API} \hl{became} ``\textit{nicely aligned with the Flutter \ac{API} and the way Dart code looks}'' [P1]. For example, the support for reactive programming using the \ac{CAMS} \texttt{events} stream fits directly into Flutter and allows for using e.g.~the \texttt{StreamBuilder} class when creating reactive \ac{UI} widgets.
In general, the ``\textit{deep integration with the Flutter ecosystem}'' was seen as an \hl{essential feature} by the programmers. Specifically, using Flutter's programming model, complying to Dart's official coding and documentation guidelines\footnote{\url{https://dart.dev/guides/language/effective-dart}}, the use of \texttt{pub.dev} for publishing packages and plugins, and support for Firebase\hl{, resulted in a streamlined developer experience.}

The design of \ac{CAMS} as a framework that plugs into the programmers' own app was also \hl{perceived as useful}. 

\begin{quote}
\emph{
I've previously done another app using another sensing framework. Compared to this, the use of \ac{CAMS} is much better. You just add the \ac{CAMS} flutter plugin and start using it -- the design of your app is independent of this. 
}
[P1]
\end{quote}

The same programmer also \hl{provided feedback on} the ability to \hl{consume} \ac{CAMS} data sampling events in the app (by just listening in on the study controller's \texttt{events} stream), \hl{since he found this to be an essential} part of \ac{CAMS}; ``\textit{... in the other framework, this was not possible at all}''.

The support for extending \ac{CAMS} using the sampling package and data manager services was also considered an \hl{essential feature} of the framework. 

\begin{quote}
\emph{
We did spend quite some time implementing the Movisens Flutter plugin -- we had a lot of issues with undocumented behavior in the Movisens Android \ac{API}. However, once we had the plugin ready, creating a \ac{CAMS} sampling package was straightforward. And when adding the Movisens measure to our app, we saw Movisens data floating in immediately.
}
[P4]
\end{quote}


\hl{The modular design of sampling packages that are loaded on compile time and only included if needed, is a feature implemented as a direct result of programmer feedback.
This ability was designed together with the MUBS programmers} when Google announced that they are limiting which apps are allowed to ask for permission to access the phone call and SMS logs\footnote{\url{https://android-developers.googleblog.com/2018/10/providing-safe-and-secure-experience.html}}. 
\hl{Once the \texttt{phone\_log} and \texttt{text\_message\_log} measures were moved to} the \texttt{communication} sampling package, \hl{and MUBS did not include this package,} the app \hl{was} approved by Google and released in the Play Store.



%


\hl{During the use of \ac{CAMS},} the programmers also \hl{helped identify} limitations \hl{of} the framework. 
Firstly, the benefit of Flutter being cross-platform also comes at a cost in terms of not being able to directly access native \ac{OS} \acp{API}. For example, the Flutter plugin for sensor access has a more limited \ac{API} compared to e.g.~the native Android sensors \ac{API}. 
Secondly, some programmers argued that \ac{CAMS} comes with a fairly complex study domain model including the study, trigger, task, and measure classes (see Fig.~\ref{fig:carp_domain}). 
However, programmers also liked the flexible and expressive manner a study could be set up,
so finding an appropriate balance was important. 
The sampling schema model was  \hl{introduced} to address part of this complexity.
%
%
Thirdly, the programmers highlighted that there is a huge difference between \textit{using} \ac{CAMS} \hl{and its built-in features} and \textit{extending} it with new functionality. Especially the documentation and code examples should to a much larger extent reflect the difference between \hl{using versus extending} the framework.
\hl{This feedback from the programmers has been incorporated in the latest release of the framework and its online documentation (see Appendix~\ref{app:online_resources})
}

\begin{table}[t]
\caption{\hl{Overall statistics on the \acf{CAMS} apps: MUBS and mCardia.}}
\label{tab:benchmark}
\small
\centering
\begin{tabular}{p{60 mm} r r }
\toprule
\textbf{} & \textbf{MUBS} & \textbf{mCardia} \\

\midrule
Total number of users                           & $174$	    & $6$                       \\	
Total days running                              & $305$	& $83$                      \\	
Total no.~ of data points                       & $588,434$	& $2,021,303$               \\	
Data points per user (avg. $\pm$ SD)        & $3,375 \pm 10,613$ & $336,884 \pm 302,427$    \\	
Data points per day (avg. $\pm$ SD)         & $1,928 \pm 2,083$ & $24,353 \pm 11,509$                \\
Data points per day per user (avg. $\pm$ SD) & $129 \pm 229$	& $14,632 \pm 2,82$          \\	
\bottomrule
\end{tabular}
\end{table}

\section{Discussion}

\subsection{Meeting the Design Goals}

The initial design goals of the \ac{CAMS} framework were to provide a reactive, unified programming \ac{API} with a cross-platform, extensible, and maintainable runtime architecture. 
%
\ac{CAMS} provides the programmer with an expressive, yet simple \ac{API} for configuring a study, executing it via a study controller, and consuming the incoming data as a stream of events in the app which can be mapped, filtered, transformed, saved, and uploaded as specified in configured plugins.
Sampling can be adapted and controlled (paused/resumed) at runtime and executes in a non-blocking asynchronous manner, preventing it from interfering with the responsiveness of the app and user experience.
%
\ac{CAMS} provides a uniform programming model to enable sensing of different on-board phone sensors, e.g., location and pedometer, logs (e.g.~call log), external wearable devices (e.g.~an \ac{ECG} monitor), and online web services (e.g.~weather information). Data sampling is configured via `measures' and sampling support and data formatting is done via sampling packages. 
\hl{Data upload to different local or remote data back-ends is supported by plugable data managers.}
By leveraging the Flutter and Dart technology, the app programmer is able to add data sampling that runs cross-platform on both iOS and Android using the same \ac{CAMS} \ac{API} resulting in a single codebase.


The most important design goal of \ac{CAMS} is to support extensibility and the framework has `hooks' that support adding new sampling measures (via the `sampling package' concept), adapting sampling (via the `sampling schemes' concept), transforming and protecting data in different ways (via the `data transformer' concept), and storing and uploading data in different ways (via the `data manager' concept). Hence, \ac{CAMS} allows for in-depth customization, as demonstrated by the \hl{two} quite different apps, one for mental health and one for cardio-vascular diseases.
Moreover, by leveraging the Dart/Flutter \hl{package manager ecosystem}, \ac{CAMS} is easily distributed and linked when programmers want to use the framework in their code.
This setup also allows third-party developers to distributed and share new packages for extending \ac{CAMS}, including new sampling packages and data managers.

The evaluation of \ac{CAMS} shows that using Flutter and Dart for implementing mobile sensing \hl{did not come with a penalty or overhead in memory and battery consumption} as compared to native mobile sensing apps implemented for Android. 
\hl{A study of sampling coverage on both Android and iOS revealed close to $100\%$ coverage.}
Hence, if a programmer wants to use Flutter (with all the benefits that come with it), \ac{CAMS} would be a useful framework for adding mobile and wearable sensing. The implementation and release of two real apps -- MUBS and \hl{mCardia} -- shows that \ac{CAMS} scales to non-trivial apps beyond the framework authors' own apps.
In addition, the \hl{programmer-centered design of the \ac{API} and documentation involving five} programmers \hl{helped to make the framework gradually more} useful and usable. 

\subsection{Limitations}

\ac{CAMS}, as well as the \hl{studies} presented here, also have a set of limitations.
%
First, even though Dart compiles to native iOS and Android, \hl{the} Flutter \hl{\ac{API}} is \hl{different from the} native \hl{\acp{API}},
and the underlying \ac{OS} features and service cannot be accessed directly. 
Access to e.g.~the \hl{location} \ac{API} is limited in Flutter compared to the Android sensor \ac{API}. 
\hl{This can be solved by using platform channels in Flutter which allow passing messages to/from the host platform, thereby giving access native services and \acp{API} where needed.}
This, however, implies that when creating a Flutter plugin, \ac{OS}-specific libraries for both iOS and Android need to be implemented. Hence, the uniform and unified \ac{API} of \ac{CAMS} comes with a cost of implementing the need to master three programming ecosystems: Flutter, iOS, and Android. 

Second, in terms of technology Dart/Flutter did prove very scalable and robust, as shown in the technical evaluation.
However, there are still a few technical limitations and open issues with Dart. Most importantly, there is an unsolved issue in the core Dart runtime, that an app crashes if a Dart isolate accesses a platform channel. 
This is a limitation for the implementation of \ac{CAMS} since data sampling almost always relies on a platform channel to access data from the native \ac{OS} and therefore isolates cannot be used for data sampling.
However, this did not seem to be a problem during our technical evaluation and we have not experienced any problems in terms of robustness during the deployment of real-world apps.
But -- support for runtime isolation of data sampling probes is planned to be implemented, once this issue is fixed by the Dart team. 

Third, even though Dart/Flutter is one of the rising technologies for app development, widespread adoption is still limited. Therefore, user \hl{involvement in the design} of \ac{CAMS} was limited to five programmers. Even though these were experienced programmers developing real apps for release, the current insight is limited to the involvement of these programmers. 
\hl{Further input from other programmers in the continued enhancement of \ac{CAMS} will be needed.}

\section{Conclusion}
\label{sec:conclusion}

%
\acf{CAMS} is a cross-platform (Android/iOS) mobile programming framework, which in addition to state-of-the-art mobile and wearable sensing provides a modern reactive programming \ac{API} with a unified approach to data sampling, management, transformation, usage, storage, and upload across different types of data sources and data storage facilities. \ac{CAMS} provides support for data sampling from on-board mobile sensors (e.g.~accelerometer, location, and step counter), from phone logs (e.g.~call log), from off-board wearable sensors (e.g.~\ac{ECG} monitor), and web-based services (e.g.~weather forecast).
All currently supported measures are listed in Table~\ref{tab:measures} and new measures are continuously being added in the form of released sampling packages.
\ac{CAMS} provides a stream-based \ac{API} which allows for flexible stream operations, including data transformation and privacy-preserving functions.
Most importantly, however, as a programming framework, \ac{CAMS} is designed to be highly extensible.
Many framework `hooks' for further extension are available, including support for adding new data sampling methods, data transformations (including privacy functions), data storage, and upload to cloud-based servers. As a publicly available framework for Flutter, \ac{CAMS} leverages the Dart package manager ecosystem for easy code releases and sharing of $3^{rd}$-party extensions.
On a more general level, the \ac{CAMS} architecture shows how a framework for mobile and wearable sensing can be designed to be extensible and flexible for different data sampling needs, data formats, data transformation, data management, and different application use cases.

\hl{The design of \ac{CAMS} was informed by a team of programmers and their continuous feedback on the usefulness and usability improved the \ac{API} and online documentation.} In addition, \ac{CAMS} has been thoroughly evaluated, both technically and in terms of usefulness in \ac{mHealth} application development.
%
%
%
Even though the use of \ac{CAMS} is still in its early stage -- programming frameworks are typically judged by their long-term use -- the present results show that \ac{CAMS} performs similar to native sensing platforms, could be used to build and release two \hl{non-trivial and} very different \ac{mHealth} applications, and that programmers found the framework useful and usable.
\hl{As such, we conclude that \ac{CAMS} might be a good choice for adding mobile and wearable sensing to cross-platform Flutter apps, using either the default-provided sensing capabilities, or when needing to extend it with support for custom measures and sensors.}

The \ac{CAMS} framework has been released along with detailed online documentation on how to use the framework for adding mobile and wearable sensing to a Flutter \ac{mHealth} applications, including how to extend and customize \ac{CAMS} for application-specific data sampling, management, use, transformation, storage, and upload. 
\hl{See Appendix~\ref{app:online_resources} for details.}
We hope that others can benefit from using \hl{and extending} \ac{CAMS}.

%
\begin{acks}
The author would like to thank Steven Jeuris for detailed discussions and input on the domain model as well as thorough review of the final paper; 
Thomas Nilsson for helping implementing some of the Flutter plugins; 
and Devender Kumar for implementing the MoviSens sampling package. 
Simon Terkelsen, Darius Rohani, and Andrea Lopategui  provided valuable insight in the inner working of the existing frameworks. 
This work is supported by \grantsponsor{}{Copenhagen Center for Health Technology}{http://www.cachet.dk/} and the \grantsponsor{}{Innovation Foundation of Denmark}{https://innovationsfonden.dk/da} under the projects: ~\grantnum[http://www.cachet.dk/research/research-projects/REAFEL]{}{REAFEL: Reaching the Frail Elderly Patient for optimizing diagnosis of atrial fibrillation} and~\grantnum[http://www.cachet.dk/research/research-projects/bhrp]{}{BHRP: Biometric Healthcare Research Platform for research in psychiatric and neurological diseases using sensor technologies}.
\end{acks}

%
\bibliographystyle{ACM-Reference-Format}
\bibliography{carp_mobile_sensing}

%
\appendix



\newpage
\section{Online Resources}
\label{app:online_resources}

\acf{CAMS} and its associated sampling and backend packages have been released as Dart packages with online \ac{API} documentation. An overview of all software packages is available online at \url{https://github.com/cph-cachet/carp.sensing-flutter} as well as shown in Table~\ref{tab:resources} (at the time of writing). 

The \acf{CAMS} online tutorial on how to use the framework in a Flutter app -- including using the different packages -- is available at \url{https://github.com/cph-cachet/carp.sensing-flutter/wiki}.

Note that these online resources are constantly updated to reflect new additions and enhancements to \ac{CAMS}.

\begin{table}[h]
\caption{\acf{CAMS} software components.}
\label{tab:resources}
\small
\centering
\begin{tabular}{p{30 mm} p{60mm} p{40mm}}
\toprule
\textbf{Component} & \textbf{Description} & \textbf{url} \\

\midrule

\textbf{Core}	& & \\	
\texttt{carp\_mobile\_sensing} & \acf{CAMS} in Flutter & \url{https://pub.dev/packages/carp_mobile_sensing}\\

\textbf{Sampling Packages} &  & \\
\texttt{carp\_communication\_ package} & 	Communication sampling package (phone, sms)	& \url{https://pub.dev/packages/carp_communication_package}\\
\texttt{carp\_context\_package} & 	Context sampling package (location, activity, weather)	& \url{https://pub.dev/packages/carp_context_package}\\
\texttt{carp\_audio\_package} & 	Audio sampling package (audio, noise)	& \url{https://pub.dev/packages/carp_audio_package}\\
\texttt{carp\_movisens\_ package} & 	Movisens Move \& ECG sampling package (movement, MET-level, ECG)	& \url{https://pub.dev/packages/carp_movisens_package}\\
\texttt{carp\_esense\_package} & 	Sampling package for the eSense ear plug device (\ac{IMU} \& button)	& \url{https://pub.dev/packages/carp_esense_package}\\

\textbf{Backends} &		& \\
\texttt{carp\_backend} &	Support for uploading data to a CARP data backend as JSON.	& \url{https://pub.dev/packages/carp_backend}\\
\texttt{carp\_firebase\_ backend} &	Support for uploading data to Firebase as both zipped files and JSON data	& \url{https://pub.dev/packages/carp_firebase_backend}\\

\bottomrule
\end{tabular}
\end{table}

\section{\ac{CAMS} Measures}
\label{app:measures}

Table~\ref{tab:measures} shows a list of currently available measures in \ac{CAMS} and which sampling package they belong to. 
Compared to other frameworks, \ac{CAMS} covers most of the common set of measures and most of them are available on both Android and iOS, which makes \ac{CAMS} a good choice for cross-platform implementation of mobile sensing. 
More packages and plugins are constantly being designed and released at \texttt{pub.dev}. 
See \url{https://pub.dev/publishers/cachet.dk/packages} for an overview of the released \ac{CAMS} sampling packages and plugins.

\begin{table}[!t]
\caption{Measures available in \ac{CAMS}, their availability on Android and iOS (+ : available, - : not available), and \hl{which} package they belong to. \hl{The top part lists} the sampling packages built into \ac{CAMS}; \hl{the} middle part lists the external sampling packages, and the lower part \hl{lists} sampling packages for wearable devices. The external packages are available for download at \texttt{pub.dev}}.
\label{tab:measures}
\small
\centering
\begin{tabular}{l c c l p{70mm}}
\toprule

\textbf{Type}  & \textbf{Android} & \textbf{iOS} & \textbf{Package} & \textbf{Description}\\

\midrule

\texttt{accelerometer}  & + & + &	\texttt{sensors}	& Accelerometer data from the built-in phone sensor \\
\texttt{gyroscope}      & + & + &	\texttt{sensors}	& Gyroscope data from the built-in phone sensor \\
\texttt{pedometer}      & + & + &	\texttt{sensors} &	Step counts from the device on-board sensor \\
\texttt{light}          & + & - &	\texttt{sensors} &	Ambient light from the phone's front light sensor \\
\texttt{device}         & + & + &	\texttt{device} &	Basic device information \\
\texttt{battery}	    & + & + &	\texttt{device} &	Battery charging status and battery level \\
\texttt{screen}         & + & - &	\texttt{device} &	Screen event (on/off/unlock) \\
\texttt{memory}         & + & - &	\texttt{device} &	Free memory \\
\texttt{connectivity}   &	+ &	+ &	\texttt{connectivity} &	Connectivity status\\
\texttt{bluetooth}      & + & + &	\texttt{connectivity} &	Scanning nearby bluetooth devices \\
\texttt{wifi}           & + & + &	\texttt{connectivity} &	SSID and BSSID from connected wifi networks \\
\texttt{apps}           & + & + &	\texttt{apps} &	List of installed apps \\
\texttt{app\_usage}     & + & - &	\texttt{apps} &	App usage over time \\

\midrule

\texttt{location}           & + & +	&	\texttt{context}	    &	Request the location of the phone.\\
\hl{\texttt{geolocation}}   & + & +	&	\texttt{context}	    &	Listens to location changes.\\
\texttt{activity}           & + & +	&	\texttt{context}	    &	Activity as recognized by OS\\
\texttt{weather}            & + & +	&	\texttt{context}	    &	Current weather and weather forecasting\\
\hl{\texttt{air\_quality}}       & + & +	&	\texttt{context}	    &	Local air quality from land-based air pollution stations\\
\texttt{geofence}           & + & +	&	\texttt{context}	    &	Entry/dwell/exit events in circular geofences\\
\texttt{audio}              & + & +	&	\texttt{audio}	        &	Records audio from the device microphone\\
\texttt{noise}              & + & +	&	\texttt{audio}	        &	Detects ambient noise from the device microphone.\\
\texttt{phone\_log}         & + & -	&	\texttt{communication}	&	Log of phone calls in/out\\
\texttt{text\_message\_log} & + & +	&	\texttt{communication}	&	Log of text messages (sms) in/out\\
\texttt{text\_message}      & + & +	&	\texttt{communication}	&	Text message (sms) events when received\\
\texttt{calendar}           & + & +	&	\texttt{communication}	&	All calendar events from all calendars on the phone\\
\texttt{survey}             & + & +	&	\texttt{survey}	        &	User surveys via the Flutter \texttt{research\_package} \\

\midrule

\texttt{movisens}           & + & -	&	\texttt{movisens}	&	\ac{ECG}-related data from the Movisens EcgMove4 device.\\
\texttt{esense\_button}     & + & +	&	\texttt{esense}	    &	Button press/release events from the eSense device.\\
\texttt{esense\_sensor}     & + & +	&	\texttt{esense}	    &	Sensor events from eSense devices.\\
\hl{\texttt{health}}        & + & +	&	\texttt{health}	    &	Wearable device data from Apple Health / Google Fit.\\

\bottomrule
\end{tabular}
\end{table}

\end{document}